\documentclass[aps,english,floatfix,twocolumn,10pt]{revtex4-1}
\usepackage[T1]{fontenc}
\usepackage{amsmath}
\usepackage{textcomp}
\usepackage[latin1]{inputenc}
\usepackage{subfigure}
\usepackage{graphicx}
\usepackage{blindtext, rotating}
\usepackage{mathtools}
\usepackage{blindtext}
\usepackage{gensymb}
\usepackage{enumitem}
\usepackage{xcolor}
\usepackage{babel}
\usepackage{hyperref}
\usepackage{braket}
\usepackage{natbib}

\usepackage{subfigure}  
\makeatother
\begin{document}


\title{Signature of half-metallicity in $\text{BiFeO}_\text{3}$}

\date{\today}
\author{Soumyasree Jena}
\email{jenasoumyasree@gmail.com}
\author{Sanchari Bhattacharya}
\affiliation{Department of Physics and Astronomy, National Institute of Technology, Rourkela, Odisha, India, 769008}
\author{Sanjoy Datta} 
\email{dattas@nitrkl.ac.in}
\affiliation{Department of Physics and Astronomy, National Institute of Technology, Rourkela, Odisha, India, 769008}
\affiliation{Center for Nanomaterials, National Institute of Technology, Rourkela, Odisha, India, 769008}

\begin{abstract}
\noindent
$\text{BiFeO}_\text{3}$ has drawn a great attention over last several decades due to its promising multiferroic 
character. In the ground state the bulk $\text{BiFeO}_\text{3}$ is found to be in the rhombohedral phase. However, 
it has been possible to stabilize $\text{BiFeO}_\text{3}$ with tetragonal structure. The importance of tetragonal phase 
is due to its much larger value of the electric polarization and the possible stabilization of ferromagnetism as in the 
rhombohedral phase. Furthermore, the tetragonal structure of $\text{BiFeO}_\text{3}$ has been reported with different 
$c/a$ ratio, opening up the possibility of a much richer set of electronic phases. In this work, we have used 
density functional theory based first-principle method to study the ferromagnetic phase of the tetragonal $\text{BiFeO}_\text{3}$ 
structure as a function of the $c/a$ ratio. We have found that as the $c/a$ ratio decreases from $1.264$ to $1.016$, the tetragonal 
$\text{BiFeO}_\text{3}$ evolve from a ferromagnetic semiconductor to a ferromagnetic metal, while passing through a 
\emph{half-metallic} phase. This evolution of the electronic properties becomes even more interesting when viewed with 
respect to the volume of each structure. The most stable half-metallic phase initially counter-intuitively evolve to 
the magnetic-semiconducting phase with a reduction in the volume, and after further reduction in the volume it finally 
becomes a metal. So far, this type of metal to insulator transition on compression was known to exist only in 
alkali metals, especially in Lithium, in heavy alkaline earth metals, and in some binary compound. 
\end{abstract}

\maketitle
\section{Introduction}
The discovery of the unusual electronic property, commonly known as half-metallic ferromagnetism, 
in Mn-based Heusler alloys \cite{groot} had started an intense research to find the existence of it 
in other materials both theoretically as well as experimentally, which continues unabated till today. 
In a half-metal, the Fermi energy is populated by electrons with only one type of spin character, either up or down. 
This unique character of the charge carriers of a half-metal makes it one of the most
suitable candidate for the spintronic devices \cite{Zutic, Coey}. 
Half-metallic ferromagnetic phase are mostly found in Heusler
alloys, and in some zinc-blende structures, colossal-magnetoresistance materials, transition metal dioxides,
double pervoskites and sulfides \cite{Katsnelson}. More recently, existence of half-metallicity has been predicted in some
Fe based perovskites such as,$(\rm{BaFeO_3})$ \cite{Zhi} and $(\rm{CaFeO_3})$ \cite{Run}. 
While the half-metallicity has been found in bulk $(\rm{BaFeO_3})$
at ambient pressure, it is found in the surface states of $(\rm{CaFeO_3})$.
\begin{figure*}[htbp!]
\centering
        \includegraphics[width=0.30\textwidth,height=0.28\textwidth]{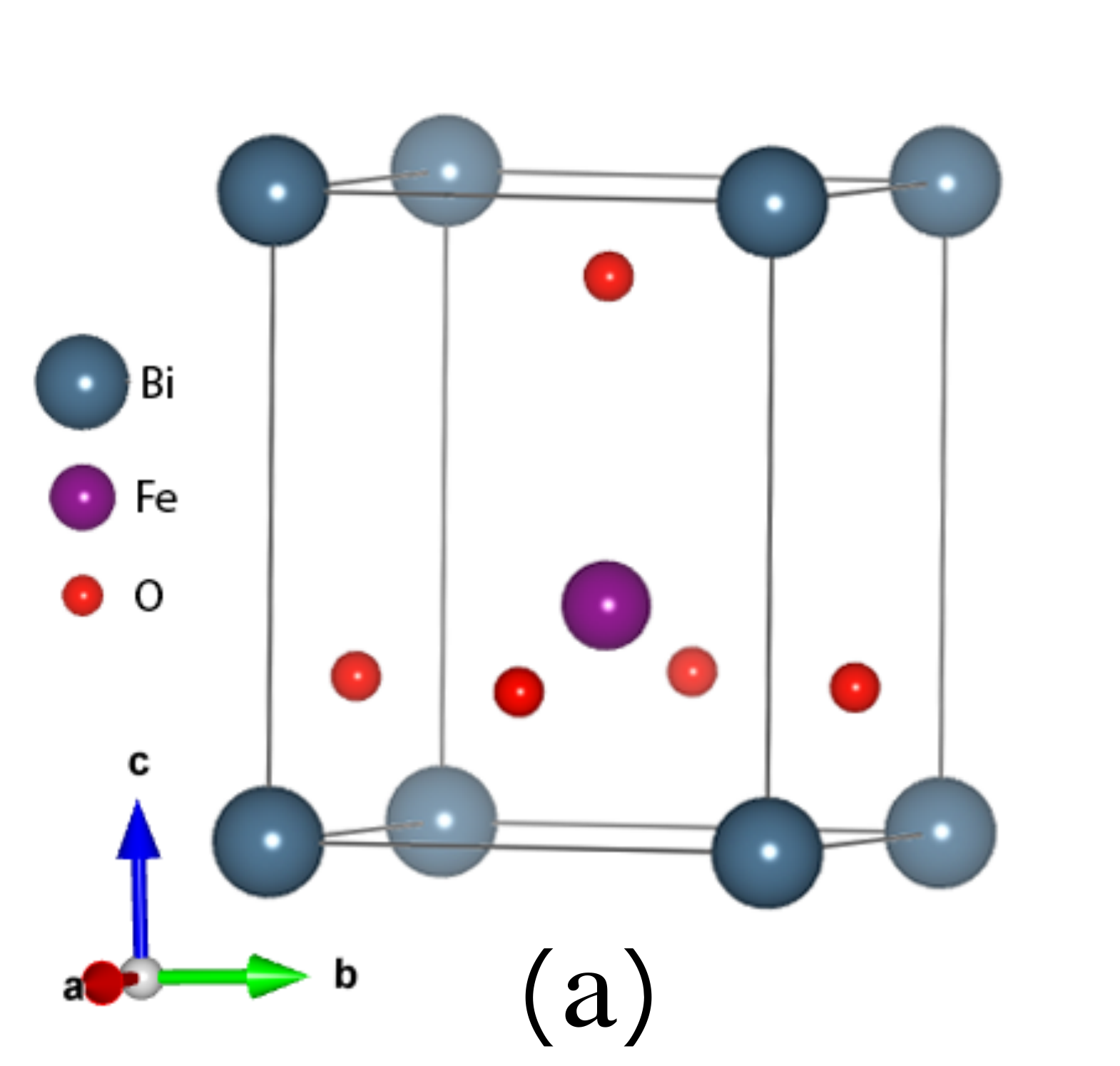}\hspace{-0.4cm}
	\includegraphics[width=0.30\textwidth,height=0.28\textwidth]{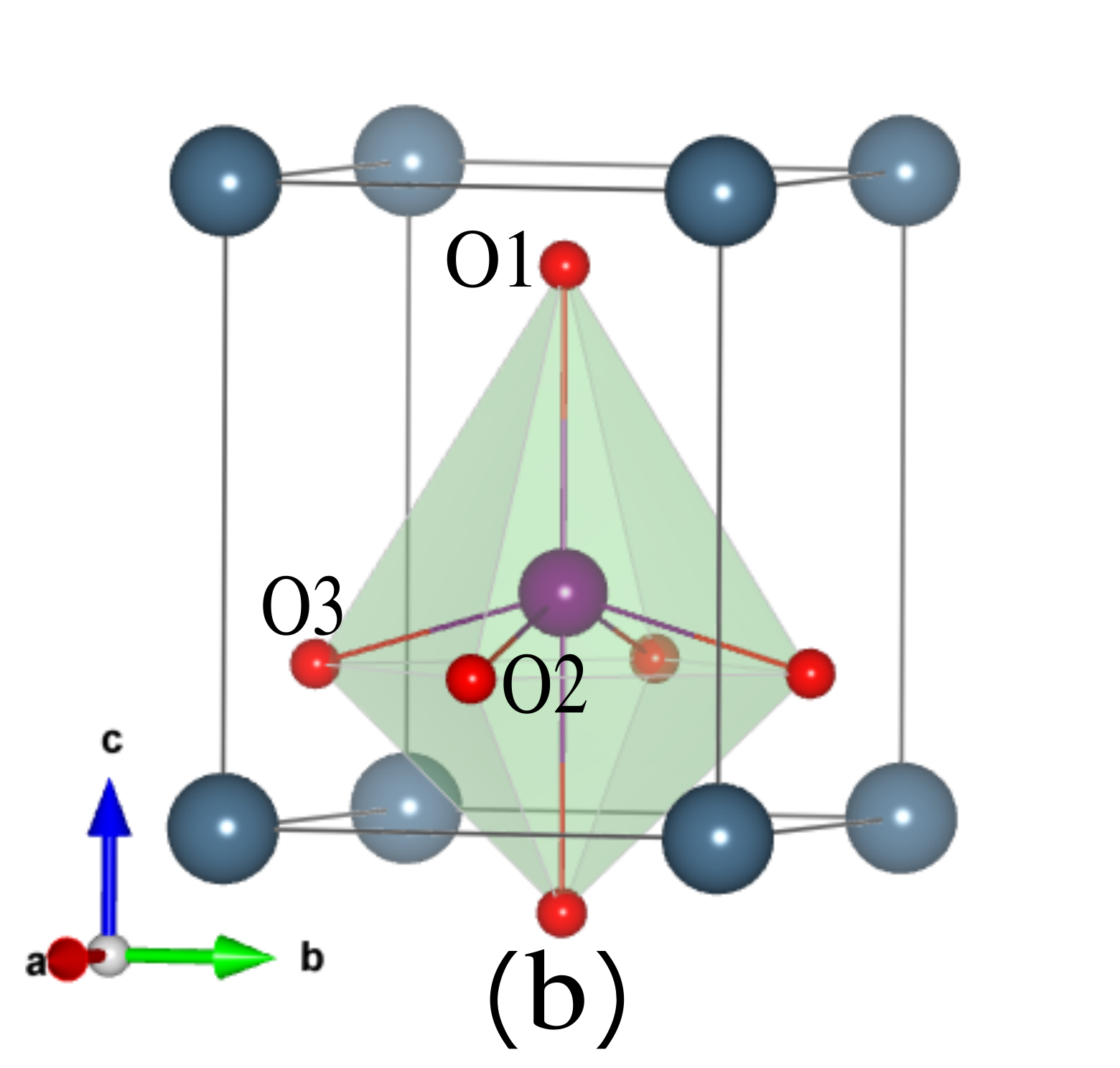}\hspace{-0.4cm}
	\includegraphics[width=0.30\textwidth,height=0.28\textwidth]{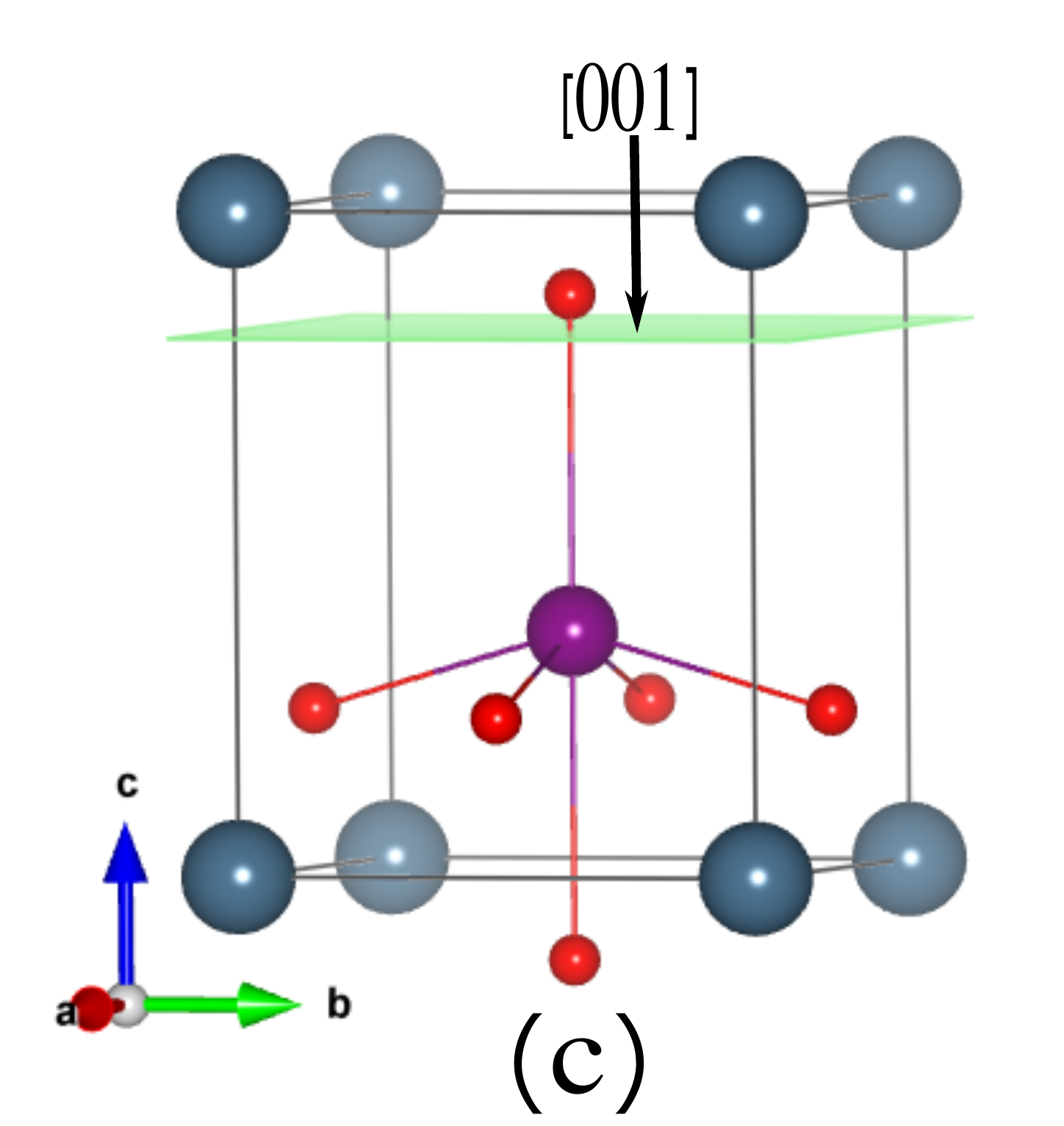}
\caption{The T-BFO cell-structure: (a) The unit-cell (UC), (b) The $\text{FeO}_\text{6}$ octahedron with one axial (O1) and 
two equatorial oxygen (O2 \& O3),(c) The displacement of axial oxygen along \{001\} plane after relaxation.}
\label{Fig:bfo-uc}
\end{figure*}
\begin{figure*}[htbp!]
\centering
        \includegraphics[width=0.28\textwidth,height=0.28\textwidth]{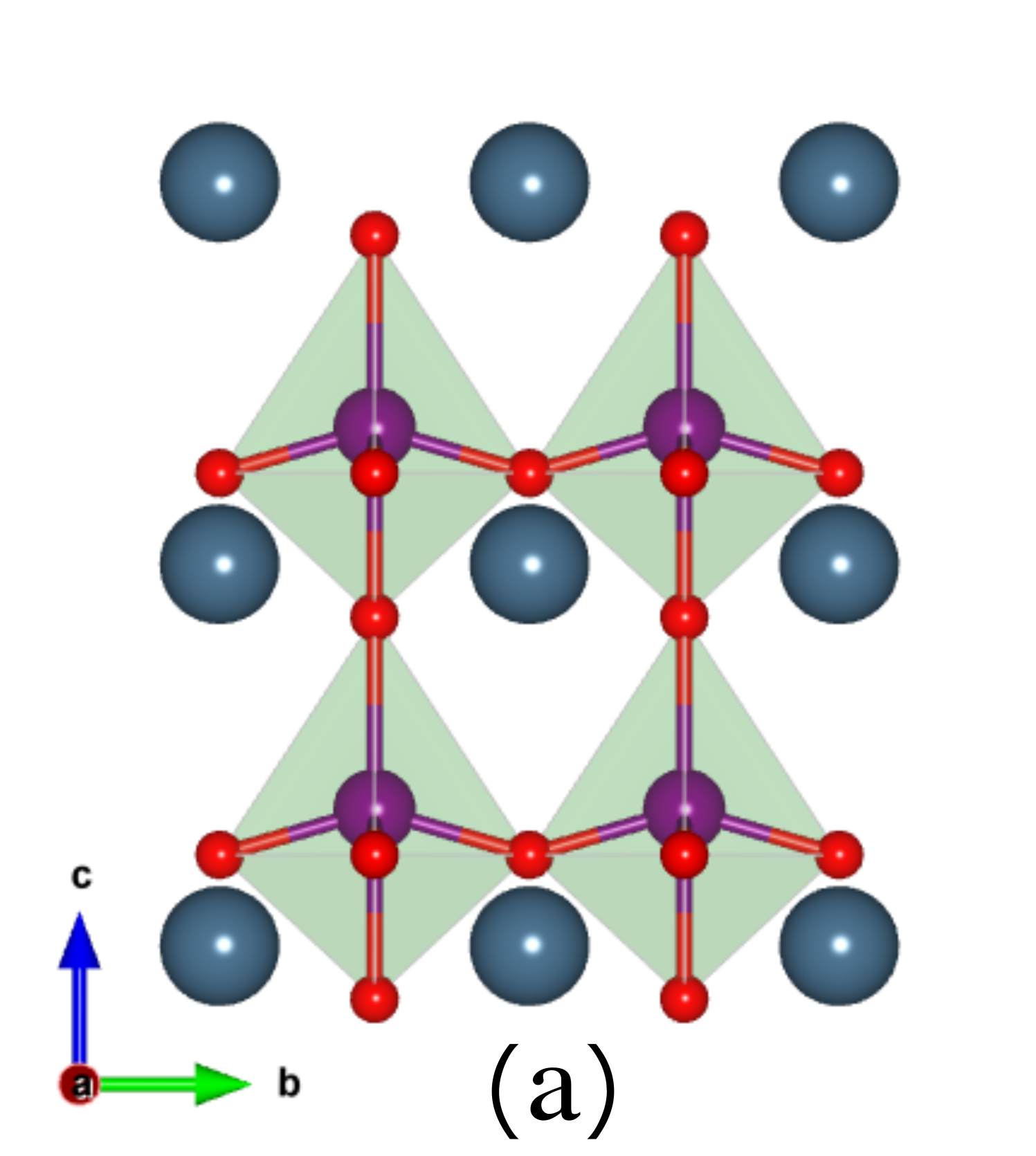}\hspace{-0.2cm}
	\includegraphics[width=0.28\textwidth,height=0.28\textwidth]{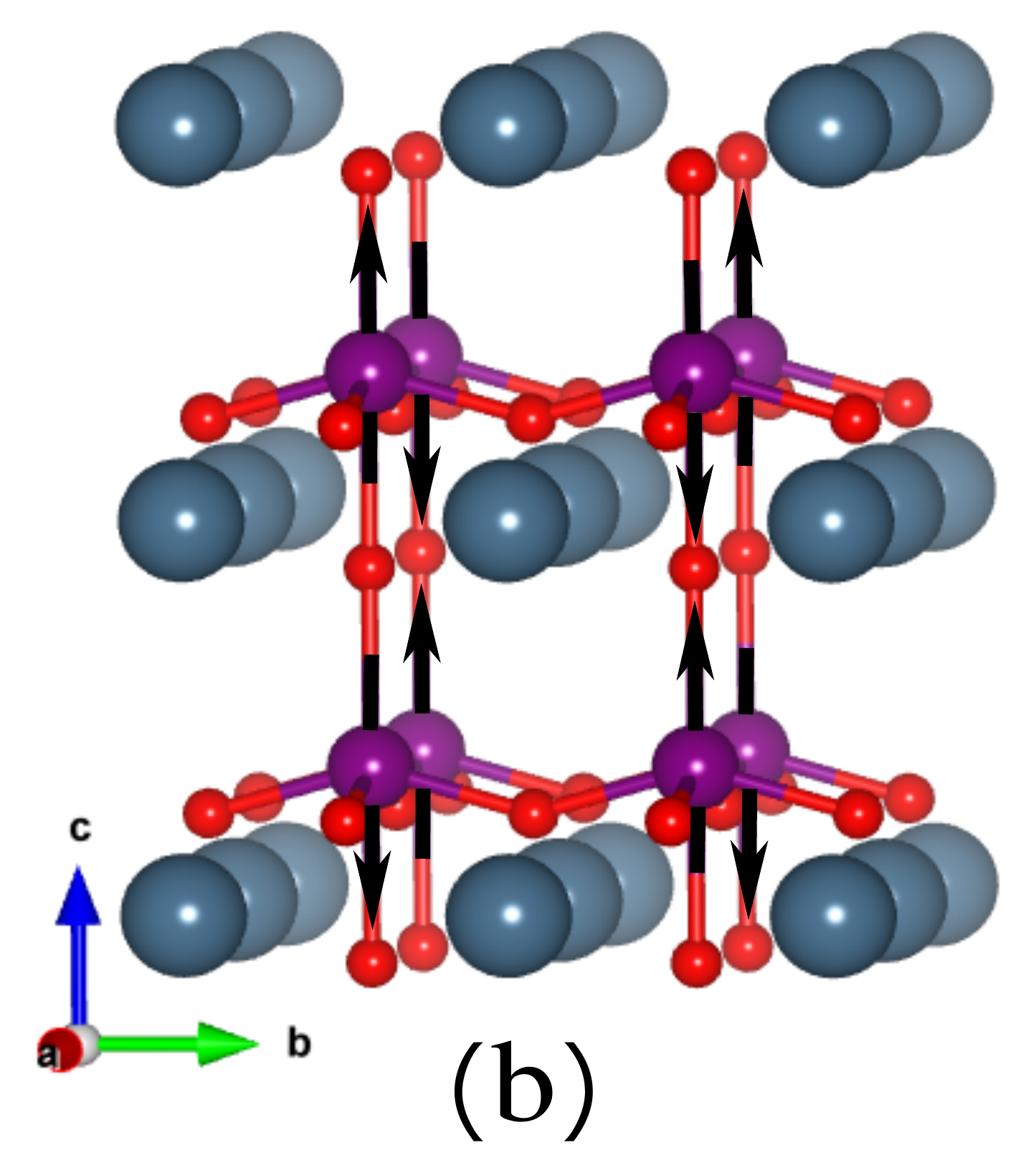}\hspace{-0.2cm}
	\includegraphics[width=0.28\textwidth,height=0.28\textwidth]{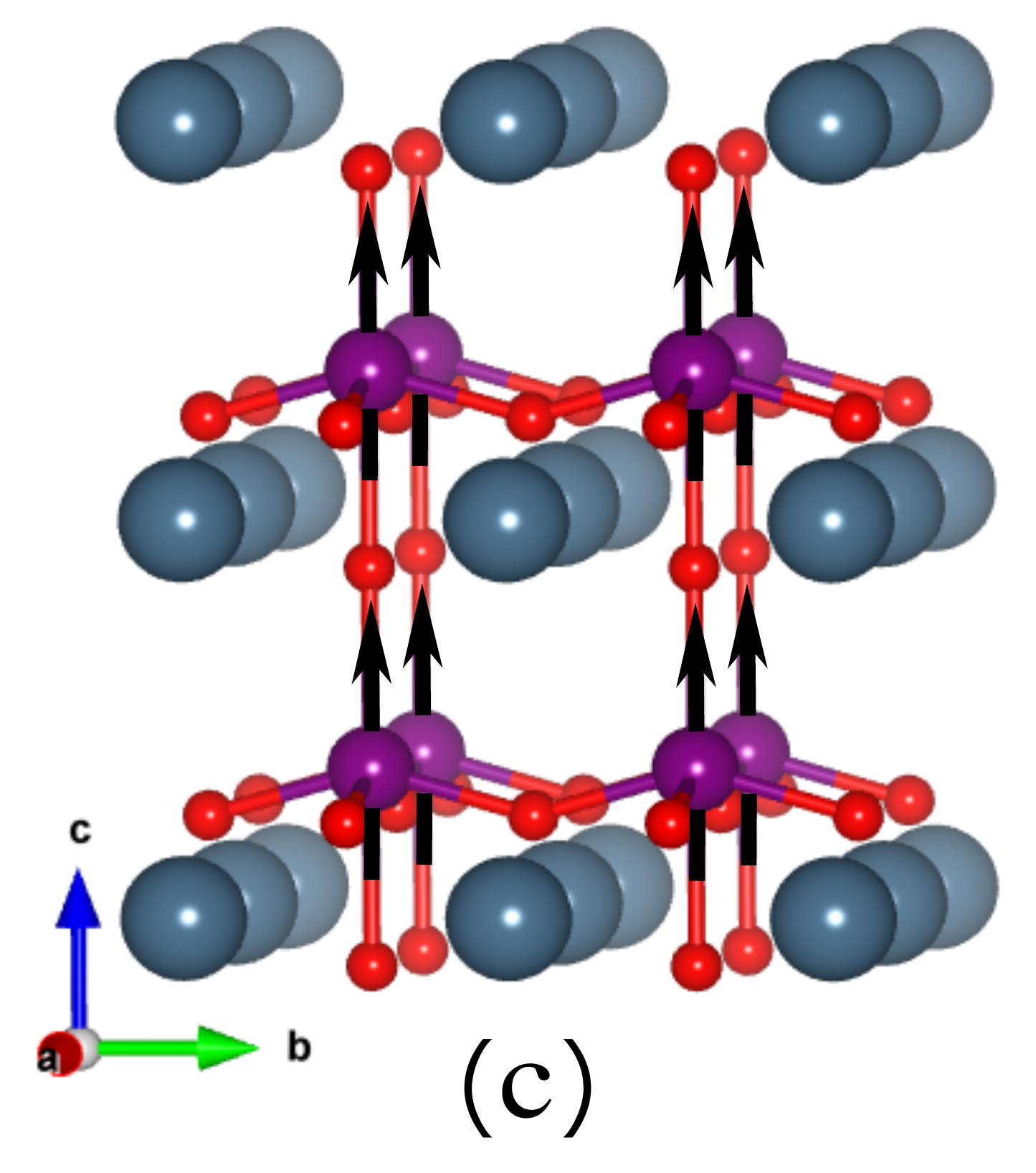}
\caption{The T-BFO supercell:(a) The $\text{FeO}_\text{6}$ octahedra with angles $\theta_1$ \& $\theta_2$, 
(b) G-type anti-ferromagnetic (G-AFM) spin imposition in $2\times2\times2$ supercell (sc),  
(c) Ferromagnetic (FM) magnetic supercell (sc), where the `$\uparrow$' and `$\downarrow$' correspond 
to spin-up and spin-down respectively. }
\label{Fig:bfo-sc}
\end{figure*}
In a parallel development, the search for a suitable
material with the possibility of external electrical control
had been quite intense for last couple of decades and multiferroic materials, especially Bismuth Ferrite ($(\rm{BiFeO_3})$),
has emerged as one of the most promising candidate in
this endeavor.
The coexistence of coupled magnetic order and electric polarization has catapulted $\rm{BiFeO_3}$ 
(BFO) as one of the leading contenders in this race. BFO has been found in various
forms having different crystal symmetries \cite{Ravindran,oswaldo}. In its ground state, bulk BFO has R3c symmetry with 
a rhombohedral perovskite structure. It has a large electric polarization along the $[111]$ 
direction of the pseudocubic unit cell along with a G-type anti-ferromagnetic ordering 
of $\rm{Fe}^{3+}$ spins. The anti-ferromagnetic ordering is superimposed with a 
long-wavelength cycloidal modulation \cite{Sosnowska-82,Sosnowska-92,R.Przenioslo,J.Herrero}. BFO 
started receiving intense  attention ever since the possibility of weak ferromagnetism (FM)
induced by the Dzyaloshinski-Moriya type interaction was predicted 
\cite{Kadomtseva-JETP2004,Spaldin-PRB2005} in this R3c structure. However, the cycloidal modulation 
forces the spins to cant which induces a spin-density-wave instead of the uniform FM state. 
A crucial step to obtain the uniform FM phase is to suppress the cycloidal modulation of the spins. There are evidences that this 
modulation can be suppressed in thin films \cite{Sosnowska-92}. A large magnetic field can achieve this 
in the bulk BFO as well \cite{Kadomtseva-JETP2004,Wardecki-JPSJ2008}.
In 2003, magnetic moments as large as $1 \mu_B /\rm{Fe}$ \cite{wang} 
was reported in heteroepitaxially constrained $\textrm{BiFeO}_3$ thin film while others 
\cite{Bellaiche, Eerenstein-Science2005,Forget-APL2005} claim to find much smaller or even non-existent 
FM moments in this system. However, there are further experimental evidences in favor of the
weak FM in BFO films \cite{Kiryukhin-PRL2011}.

Although, the existence of large ferromagnetic moment 
in the rhombohedral BFO (R-BFO) is still a debated issue, there are growing experimental 
evidences that BFO can host FM moments \cite{Ramirez-ActaMat17} and it could be quite large \cite{Anil-Kumar-EPL19}
under certain conditions. Interestingly, the heteroepitaxially constrained BFO structure belongs to 
$P4mm$ space group with a body centered tetragonal unit cell. In a subsequent first-principle based 
calculation, Ederer and Spaldin \cite{eder-spladin-prl} predicted the thin-film tetragonal (T) BFO structure 
with G-type anti-ferromagnetic (G-AFM) ordering to have much larger ferroelectric polarization as
($\approx 150 \mu C /\textrm{cm}^2$) compared to the R phase. Furthermore, using strain induced calculation, they have 
studied all the types of anti-ferromagnetic ordering in T-BFO \cite{eder-spladin-strain}. This prediction of larger 
polarization has been confirmed experimentally \cite{Zhang-Ramesh-PRL}, although the measurement of the 
polarization has not been done on pure T-BFO phase, but rather on the mixed phase of T-BFO and R-BFO.
\begin{figure*}[htbp!]
\centering
	\includegraphics[width=0.35\textwidth,height=0.32\textwidth]{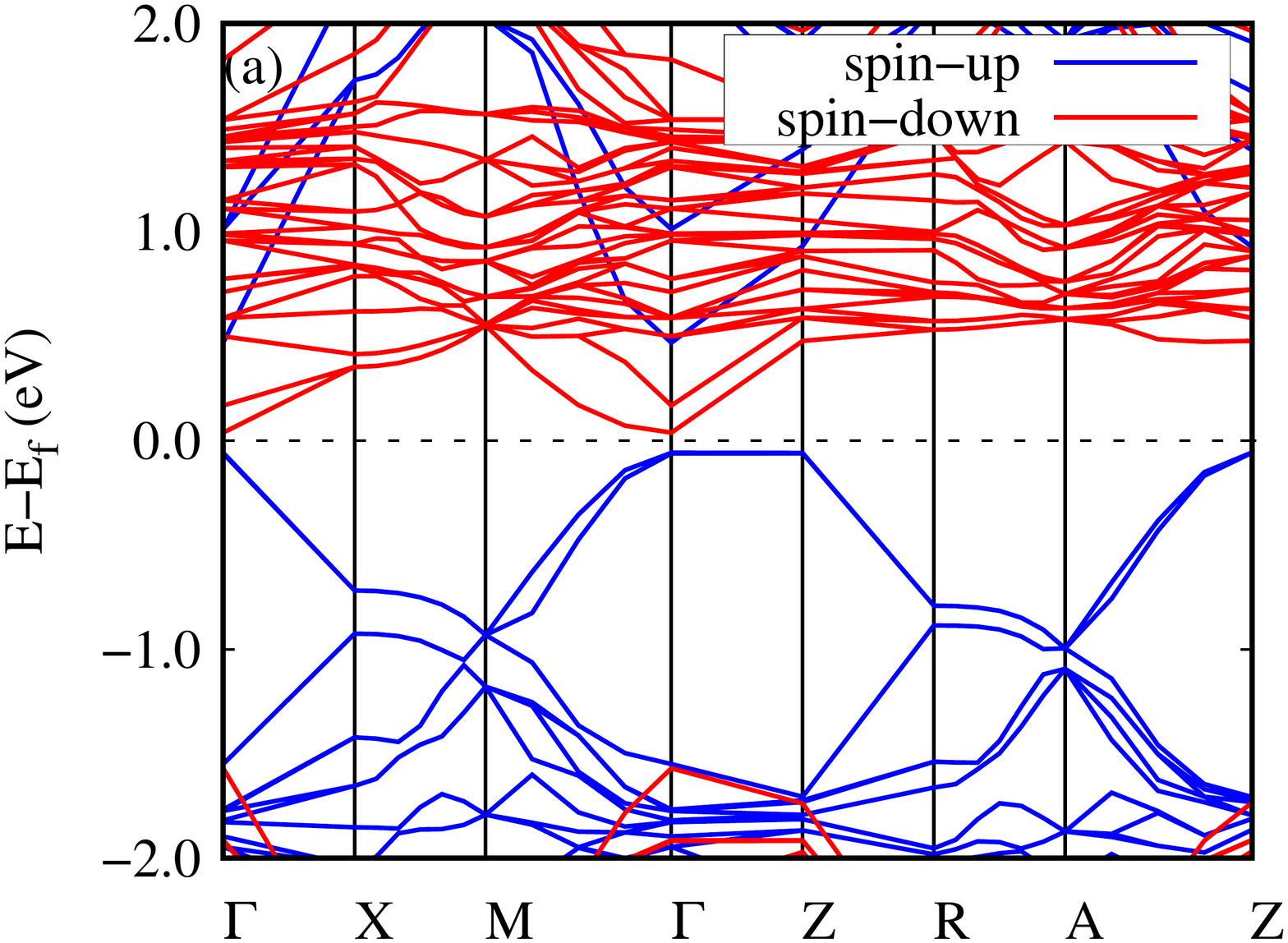}\hspace{-0.6cm}
	\includegraphics[width=0.35\textwidth,height=0.32\textwidth]{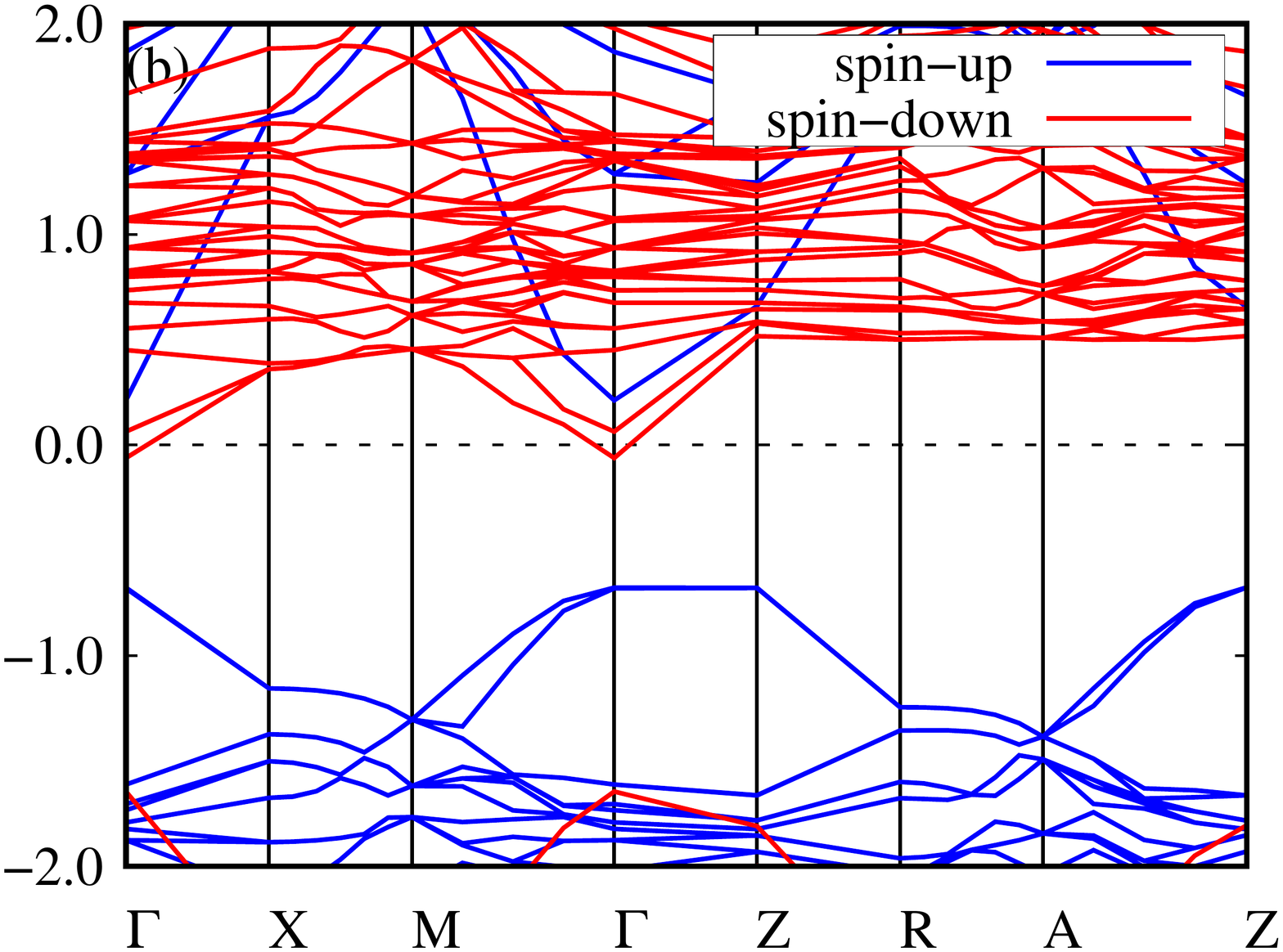}\hspace{-0.6cm}
	\includegraphics[width=0.35\textwidth,height=0.32\textwidth]{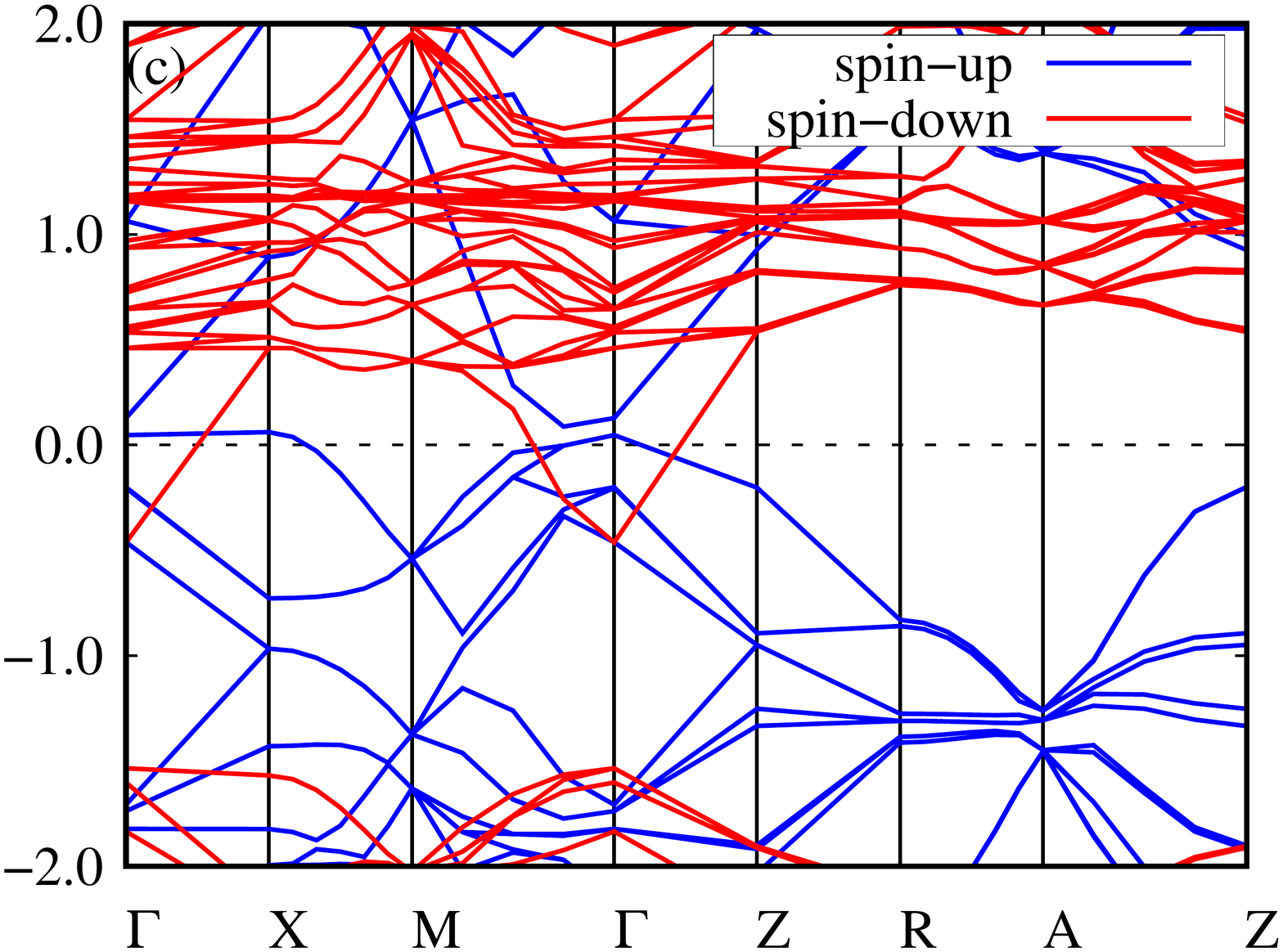}
\vspace{-0.5cm}
\caption{The Band structure (BS) with spin-up (SU) and spin-down (SD) components corresponding to the FM phase in 
(a) structure-I, (b) structure-II, and (c) structure-IV respectively.}
\label{Fig:fm-bs}
\end{figure*}

The tetragonal structure is metastable \cite{Ravindran}, with the possibility of 
a wide range of $c/a$ ratio \cite{wang, Ricinschi, Gp-Srivastava, Sando, oswaldo}. 
Major efforts have been devoted to understand and to enhance the electric polarization in 
the T-BFO structure \cite{Ricinschi}\cite{Zhang-Ramesh-PRL}\cite{eder-spladin-prl}.
However, the  issue of ferromagnetism has not been explored in pure T-BFO structure as extensively as in 
the case of R-BFO phase, although, there are density-functional theory (DFT) based first principle calculations on doped 
T-BFO systems\cite{Mn-doping, Ni-doping}. The non-uniqueness of the structural parameters of the T-BFO phase leaves the 
possibilities of exotic phases wide open. In this work, we study the ferromagnetic phase of the T-BFO structure
using DFT based first principle calculations. We have systematically studied the four different T-BFO structures (I,II,III and IV), 
having different $c/a$ ratios and which have already been reported experimentally \cite{wang, Ricinschi}. The $c/a$
ratio decreases monotonically from structure-I to structure-IV. We have found that in the presence of the uniform FM order, 
the structure-II is energetically the most stable out of all the four structures. Interestingly, this structure also hosts 
\textit{half-metallicity}, while structure-I is a magnetic semiconductor and the other two structures have metallic character. 
Additionally, we have found that the electrical properties in this system evolves counter-intuitively,
i.e., there is a transition from the half-metallic phase of structure-II to the semiconducting phase of 
structure-I along with a significant decrease in the volume.
Half-metallicity has been predicted in doped T-BFO \cite{Ni-doping} system and in some BFO-based 
heterostructures \cite{yin, nan, wei}. However, to the best of our knowledge, this is the first time the 
half-metallicity is being reported in the pure T-BFO structure. \\
The paper is organized as follows: we have first discussed the computational details in Sec.{\ref{computational-details}}, 
followed by the structural, electronic and magnetic properties in Sec.~\ref{result-discussion:structure}, 
\ref{result-discussion:electronic properties}, \ref{result-discussion:magetism} respectively. In 
Sec.~\ref{MO-CD-H}, we have discussed the role of the molecular orbitals, charge density and hybridization. Finally,
we conclude our results in Sec.~\ref{Sec:conclusion}.

\section{Computational Details}\label{computational-details}
To systematically explore the effect of the uniform ferromagnetic state in the T-BFO structure , 
we have performed DFT based first-principle calculations by the pseudopotential method as
implemented in the Quantum Espresso (QE) package \cite{qe}. The pseudopentials with Projected Augmented Wave (PAW) \cite{paw} 
type basis sets that have been used in our calculation consists of
15 valence electrons 
$(5d^{10}6s^{2}6p^{3})$ for bismuth (Bi), 16 valence electrons $(3s^{2}3p^{6}3d^{6}4s^{2})$ for iron (Fe) and 6 valence electrons 
$(2s^{2}2p^{4})$ for oxygen (O).
We have used  $GGA+U$ method with Perdew-Burke-Ernzerhof 
(PBE) \cite{pbe} type exchange 
correlation functional.  An energy cut-off of 80 Ry has been used for the subsequent calculations.
The Hubbard $U$ parameter has been taken into account for the Fe-$3d$ orbital and its strength has been set to $4.5~\rm{eV}$.
With this choice of the Hubbard $U$, the theoretical band gap has been found to be reasonably close to the 
experimentally reported value in T-BFO \cite{chen, Cameliu-expt-bg}.   
${\text{BiFeO}_\text{3}}$ crystallizes in a tetragonal phase with space group $P\it{4mm}$ (99) with different possible 
$c/a$ ratios. Since it is well known that in the ground state the magnetic moments are ordered in 
G-type anti-ferromagnetic arrangement, we have used a $2\times2\times2$ supercell (sc) composed 
of 8 formula units to compare the energies of the FM and G-AFM phases.
A $6\times6\times6$  Monkhorst-Pack k-point grids \cite{Monkhorst-pack} has been used for the self-consistent calculation. 
The Methfessel and Paxton type smearing is used with a  width of 0.01 Ry \cite{Methfessel-paxton}.
Ionic relaxation has been performed with the convergence threshold of the inter-atomic forces being  
less than $ 10^{-3}$ Ry/bohr. We have used collinear spin-polarized calculation to study the FM phase. 

\begin{figure*}[htbp!]
\centering
	\includegraphics[width=0.35\textwidth,height=0.32\textwidth]{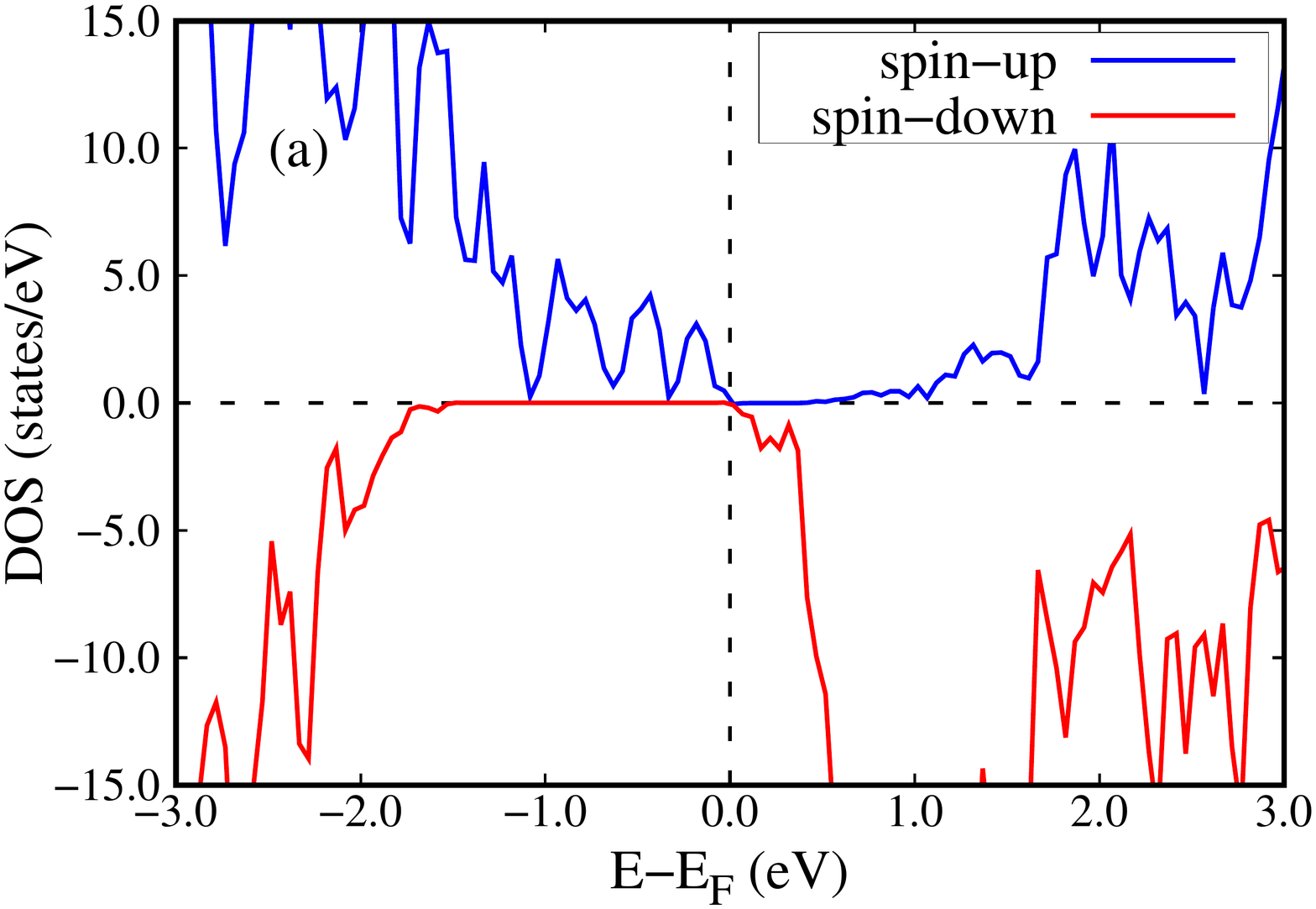}\hspace{-0.8cm}
	\includegraphics[width=0.35\textwidth,height=0.32\textwidth]{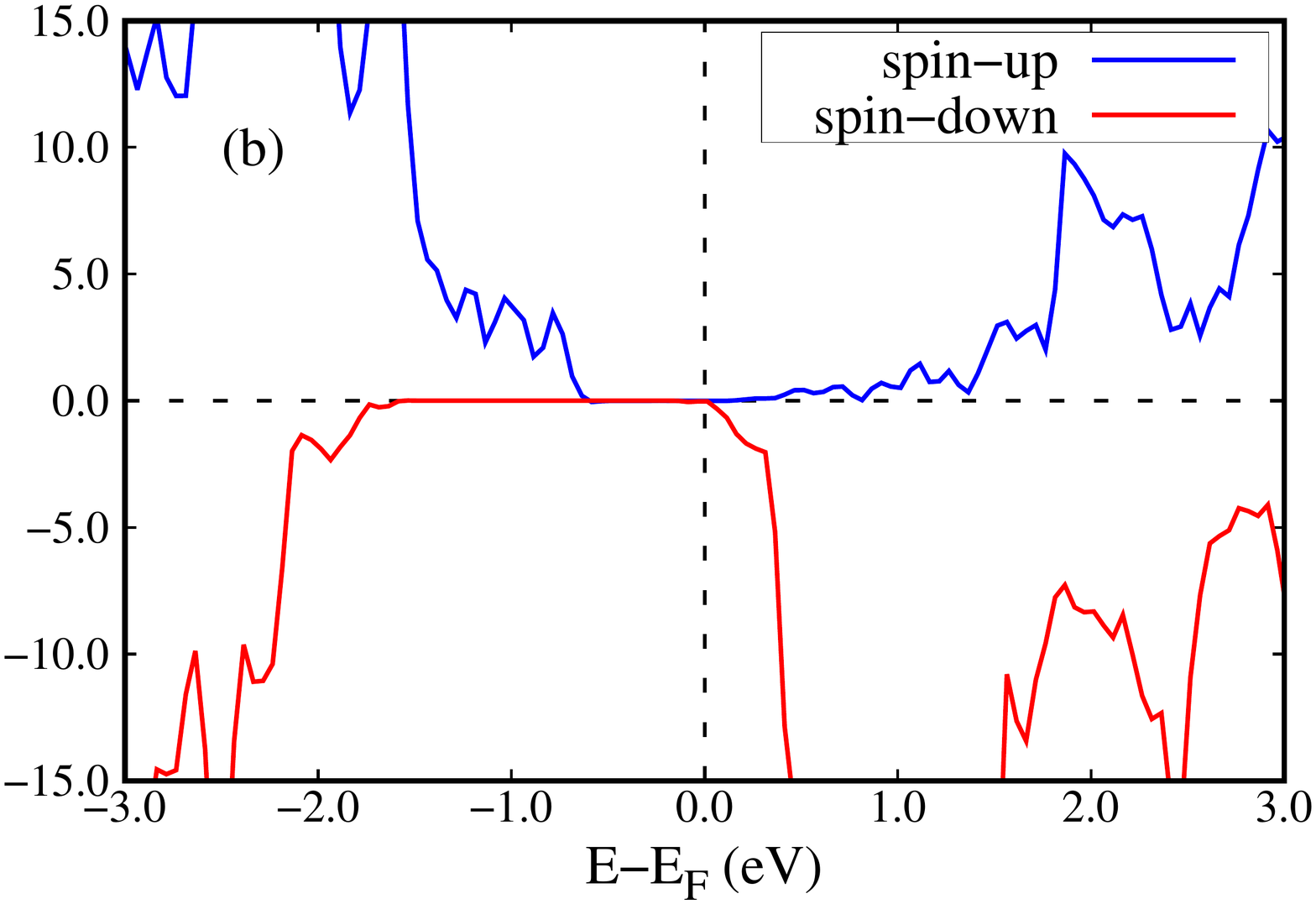}\hspace{-0.8cm}
	\includegraphics[width=0.35\textwidth,height=0.32\textwidth]{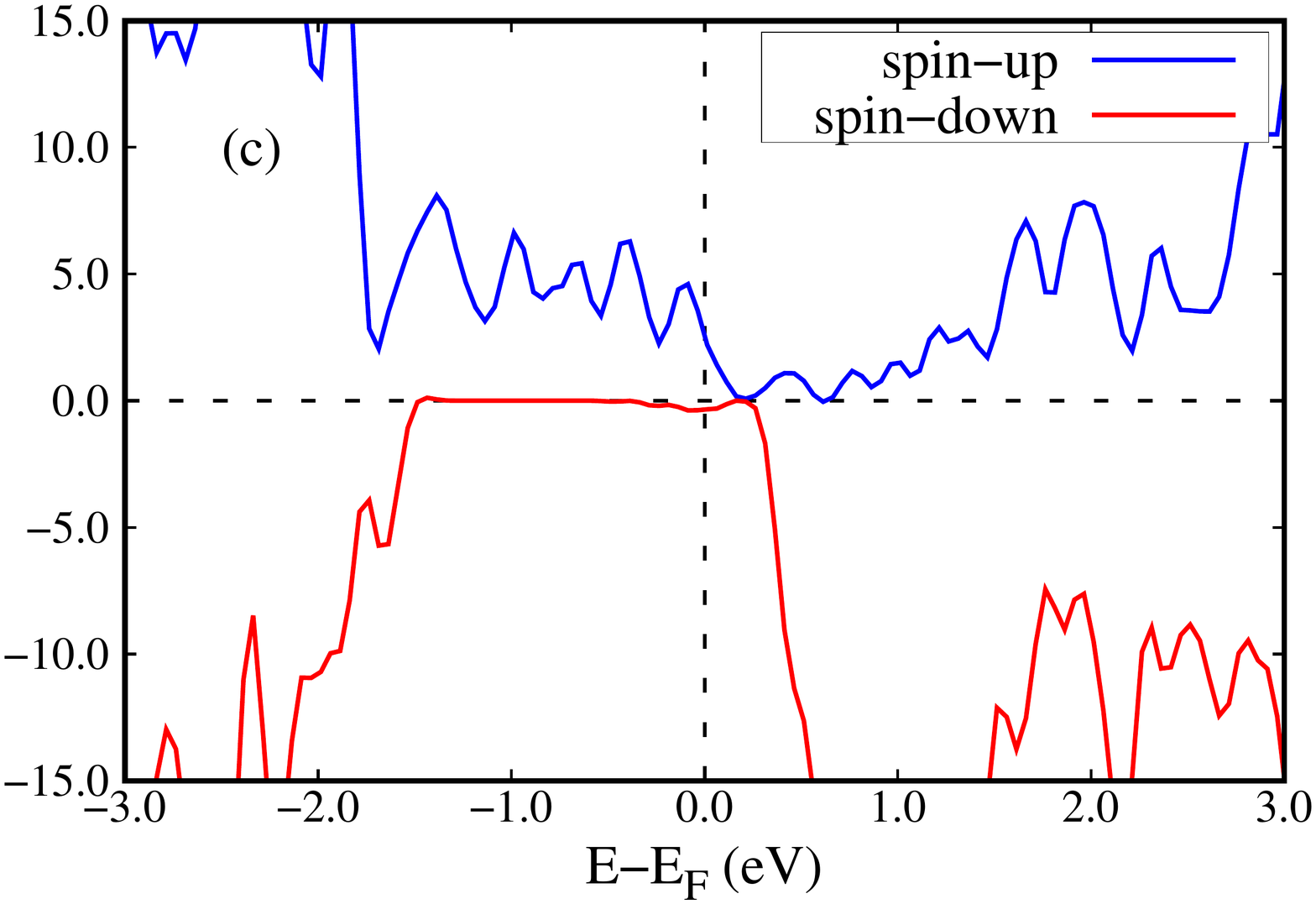}
\vspace{-0.8cm}
\caption{The total density of states (TDOS) corresponding to the FM phase in (a) structure-I, (b) structure-II, and (c) structure-IV
respectively.}
\label{Fig:fm-tdos}
\end{figure*}
\begin{figure*}[htbp!]
\centering
	\includegraphics[width=0.35\textwidth,height=0.32\textwidth]{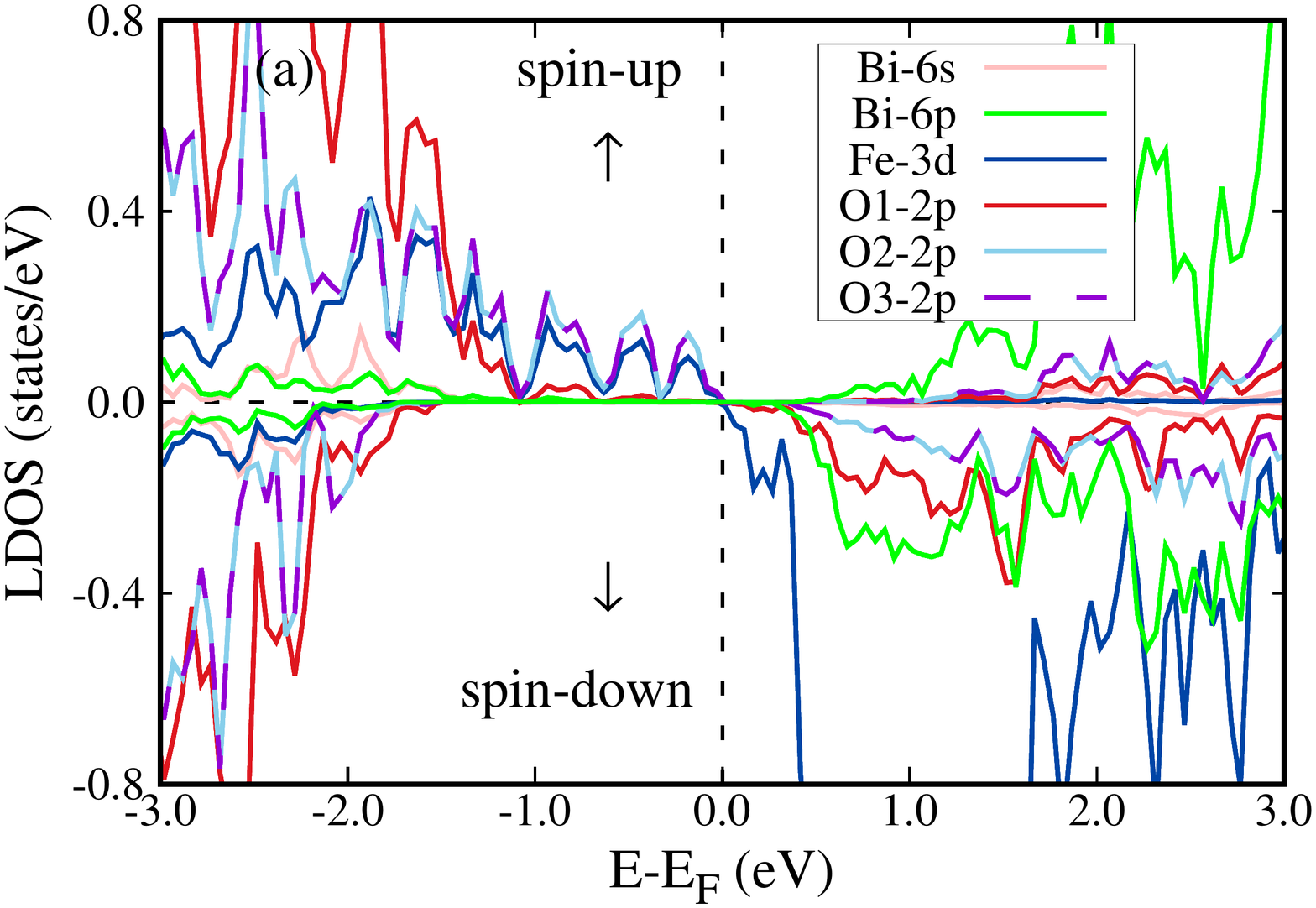}\hspace{-0.8cm}
	\includegraphics[width=0.35\textwidth,height=0.32\textwidth]{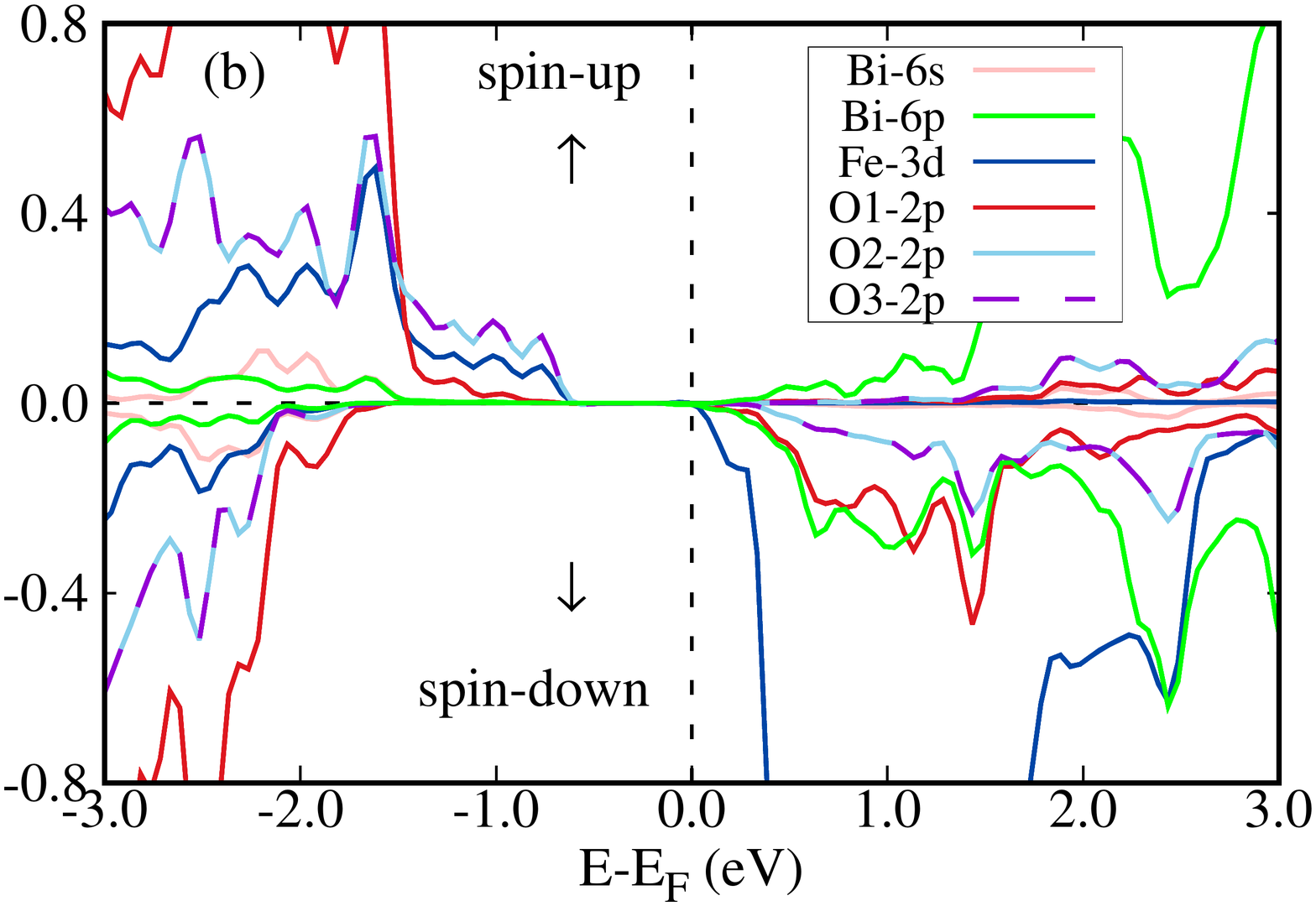}\hspace{-0.8cm}
	\includegraphics[width=0.35\textwidth,height=0.32\textwidth]{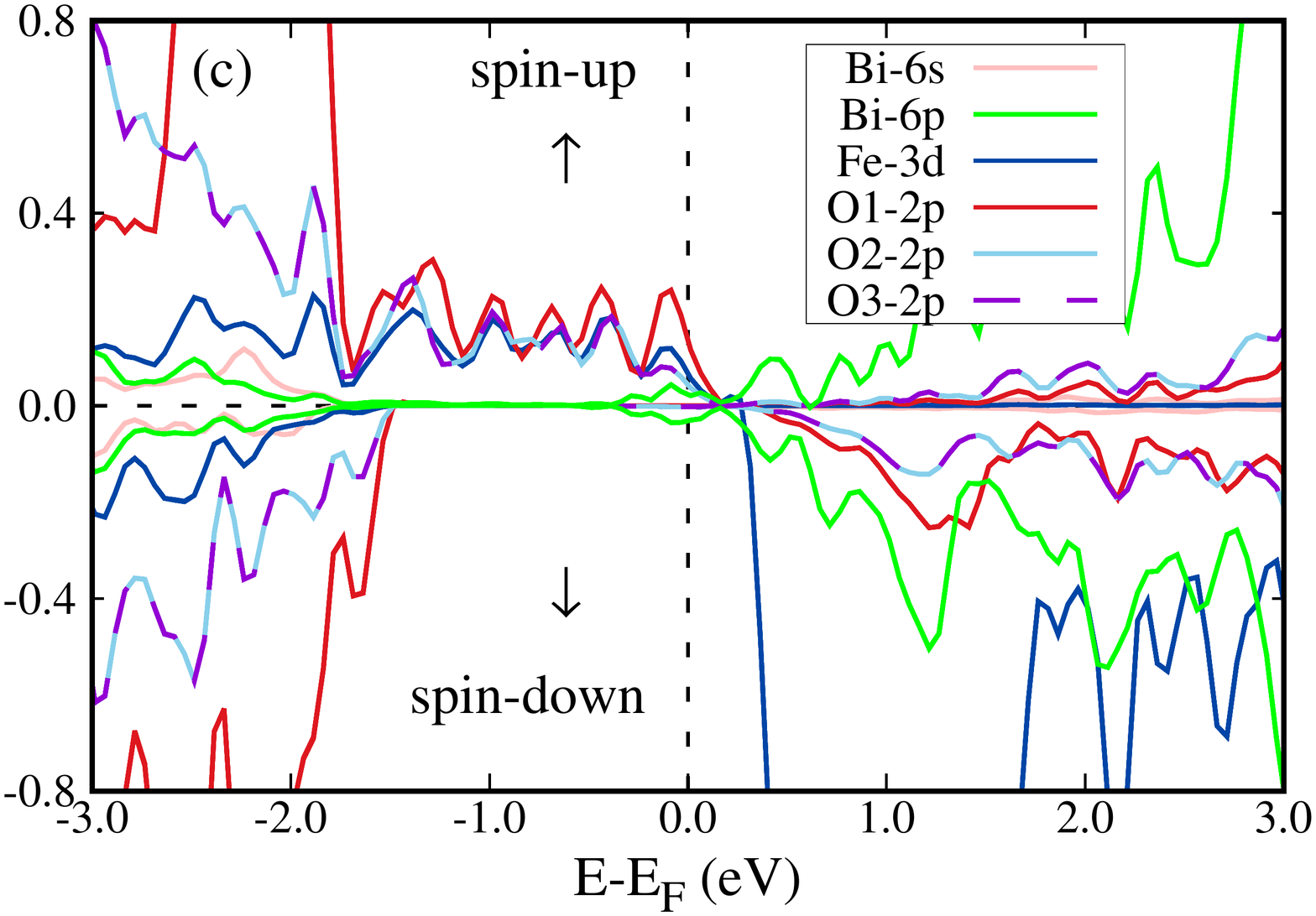}
\vspace{-0.8cm}
\caption{The local density of states (LDOS) corresponding to the FM phase in (a) structure-I, (b) structure-II, and (c) structure-IV
respectively. }
\label{Fig:fm-ldos}
\end{figure*}

\section{Results and Discussion:}\label{result-discussion}
\subsection{Structural properties}\label{result-discussion:structure}
In this work, we have used the experimentally reported parameters of the T-BFO unit-cell 
($P\it{4mm}$ space-group and $C_{4v}$ point group) \cite{wang, Ricinschi}. The highest $c/a$ ratio is 1.264 (structure- I) 
\cite{Ricinschi} and for the rest of the structures it is 1.233 (structure-II), 
1.049 (structure-III) \cite{Gp-Srivastava} 
and 1.016(structure-IV) \cite{wang}. The lattice parameter `a' of these structures are 3.67, 3.77, 3.88 and 
3.93 \AA{} respectively. 
The ionic relaxations have been performed for all the structures while keeping the lattice constant $a$ and 
the $c/a$ ratio fixed before studying the electronic and magnetic properties. The relaxations have been carried out 
after imposing the magnetic 
ordering. 

In Fig.~\ref{Fig:bfo-uc}(a), we have shown the unit-cell of the T-BFO. The $\text{FeO}_\text{6}$ 
octahedra has been shown in the Fig.~\ref{Fig:bfo-uc}(b), wherein, the axial oxygen has been named 
as `O1' while the 
two equatorial oxygen have been designated as `O2' and `O3' respectively. The lattice plane perpendicular 
to the [001] direction has been pointed out in Fig.~\ref{Fig:bfo-uc}(c). 
In Fig.~\ref{Fig:bfo-sc}(a), the supercell has been introduced with the oxygen 
octahedra.  The G-AFM and the FM type of spin-orderings, that have been used in our study, have been shown in 
Fig.~\ref{Fig:bfo-sc}(b) and (c) respectively.

In Appendix~\ref{Appendix:c/a-vs-ene}, we have presented relative energies of these four structures with 
G-AFM and FM type magnetic ordering. We have found that the G-AFM ordering is energetically more favorable than the FM phase 
irrespective of the $c/a$ ratio. However, the energy follows a similar pattern in the presence of each type of magnetic 
ordering, that is, structure-II is energetically most stable for a given type of magnetic ordering, followed by structure-I, 
structure-III and structure-IV. 
\begin{table}[htbp!]
\caption{\label{Table-1:} 
{Top row: The angle Fe-$\textrm{O}_{e}$-Fe (in unit degree, \degree), where $\textrm{O}_{e}$ is the 
equatorial oxygen O2/O3. Second row: the bond-length Fe-O1 (in unit \AA{}) Third row: Fe-$\textrm{O}_{e}$ bond-length 
(in unit \AA{}). Bottom row: the volume of the unit-cell BFO (in unit \AA{}$^3$ )}}
\begin{center}
\begin{tabular}{c c c c c c c c }
\hline
Structures &  & str-I     & str-II  & str-III & str-IV   \\
c/a ratio  &  & 1.264     & 1.233   & 1.049   & 1.016 \\
\hline
Fe-$\textrm{O}_{e}$-Fe    &  & 146.59    & 144.01  & 163.31  & 165.04 \\
O1-Fe       &  & 1.85      & 1.83    & 1.89    &  1.88  \\
Fe-$\textrm{O}_{e}$      &  & 1.92      & 1.98    & 1.96    &  1.98  \\
Volume          & & 62.48 & 66.07 & 61.27 & 61.67 \\
\hline
{\label{Table:Fe-O-Fe-bond}}
\end{tabular}
\end{center}
\end{table}
\begin{figure*}[htbp!]
\centering
	\includegraphics[width=0.32\textwidth,height=0.30\textwidth]{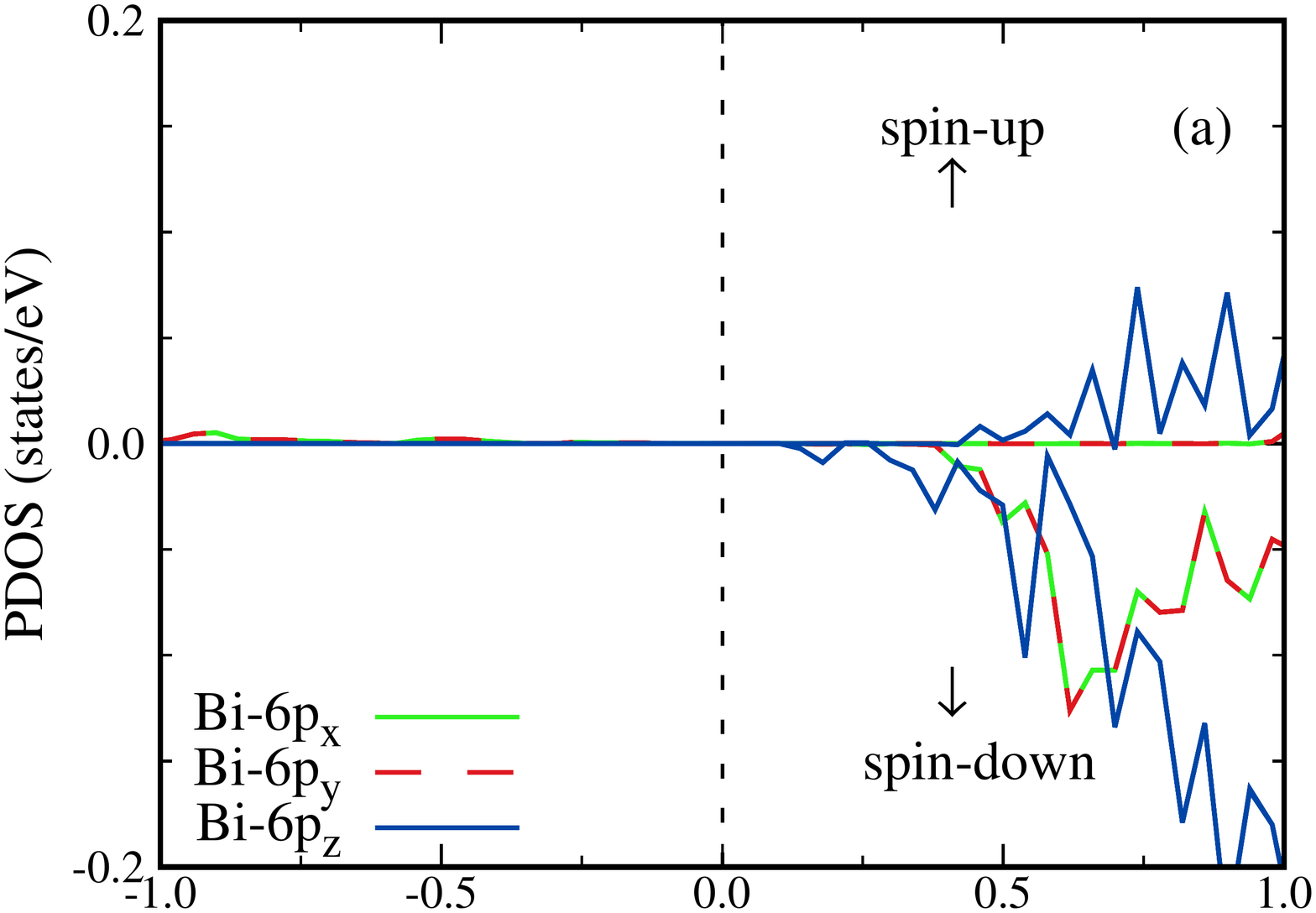}\hspace{-0.5cm}
	\includegraphics[width=0.32\textwidth,height=0.30\textwidth]{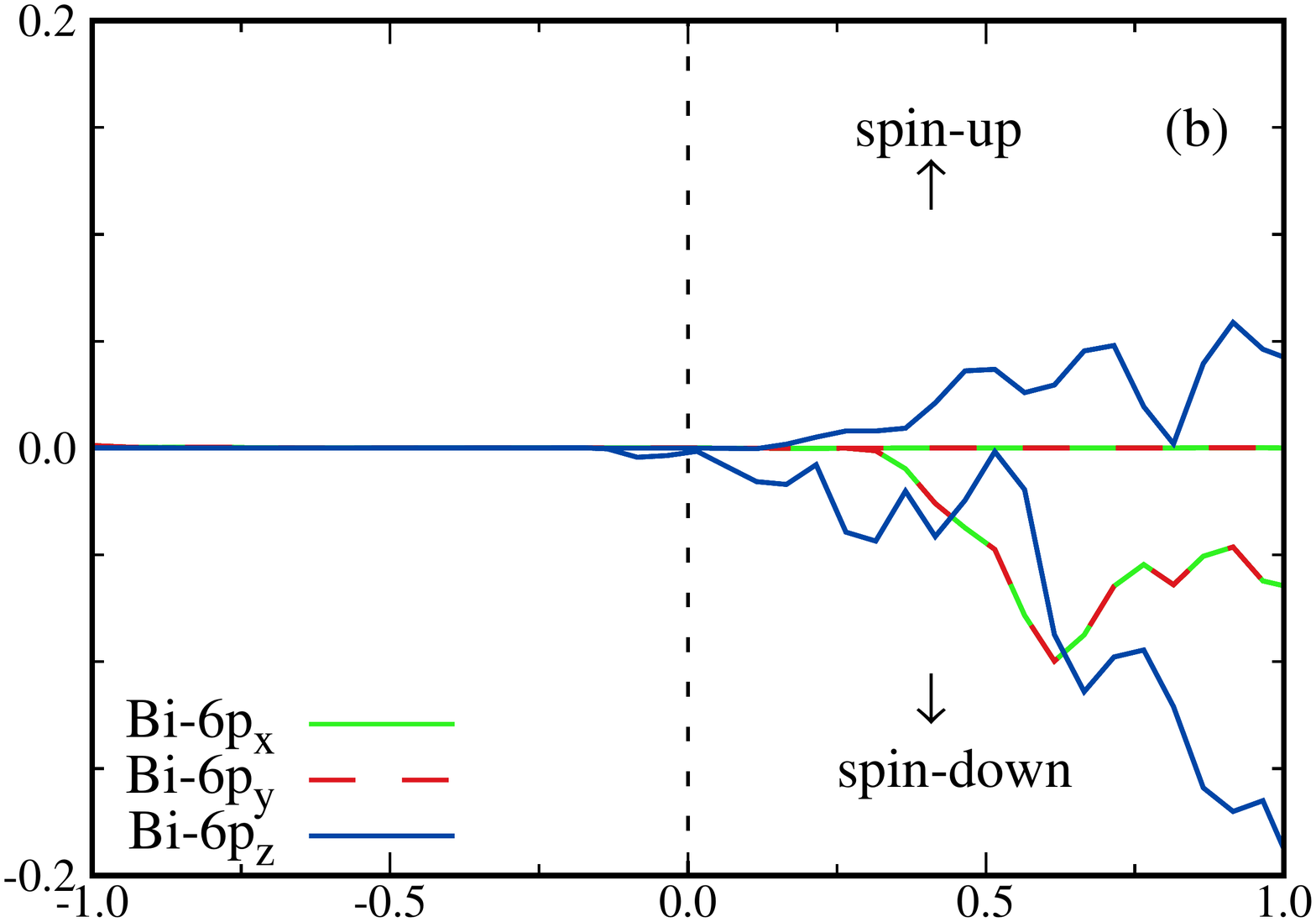}\hspace{-0.5cm}
	\includegraphics[width=0.32\textwidth,height=0.30\textwidth]{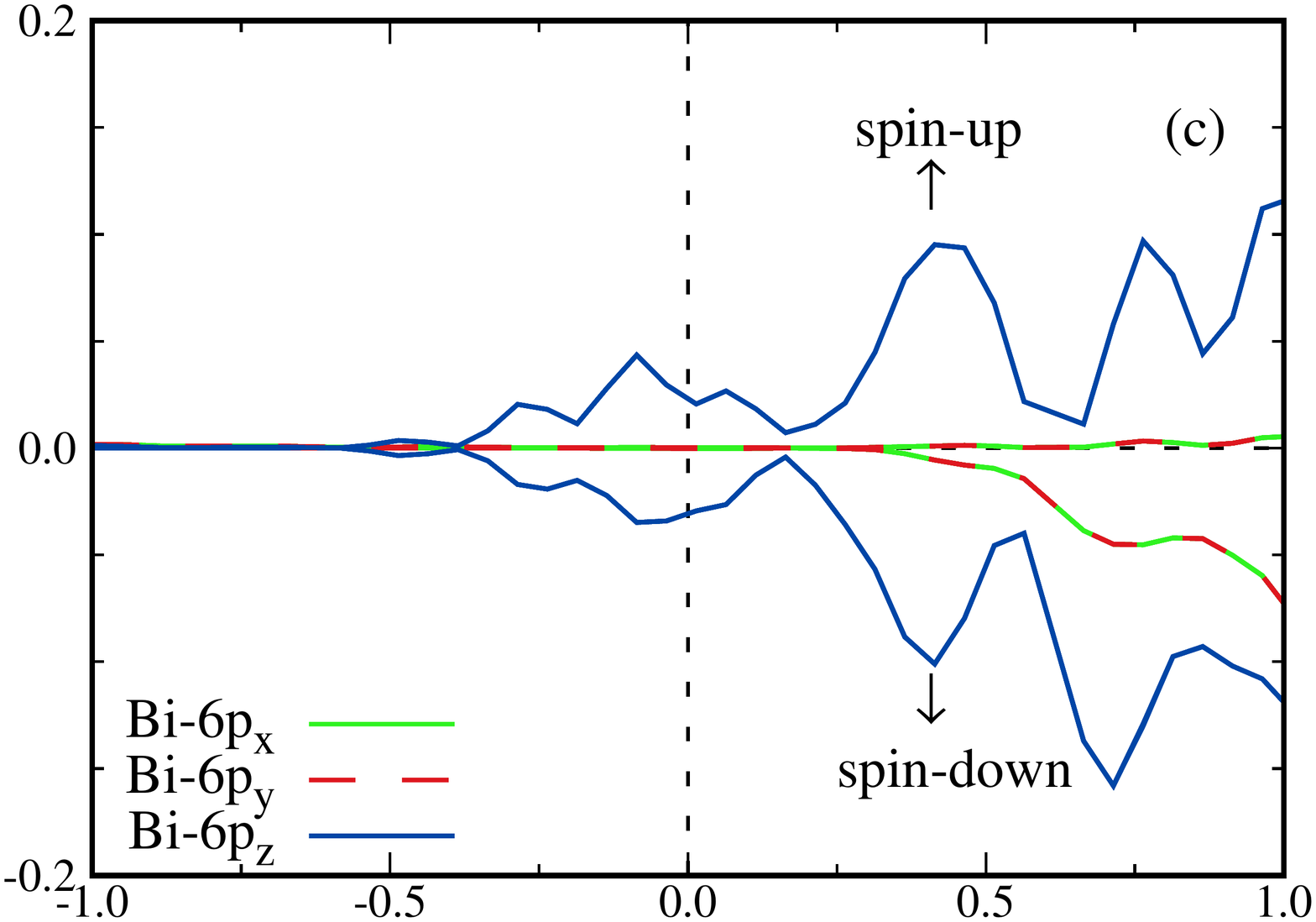}\vspace{-0.5cm}\\
	\includegraphics[width=0.32\textwidth,height=0.30\textwidth]{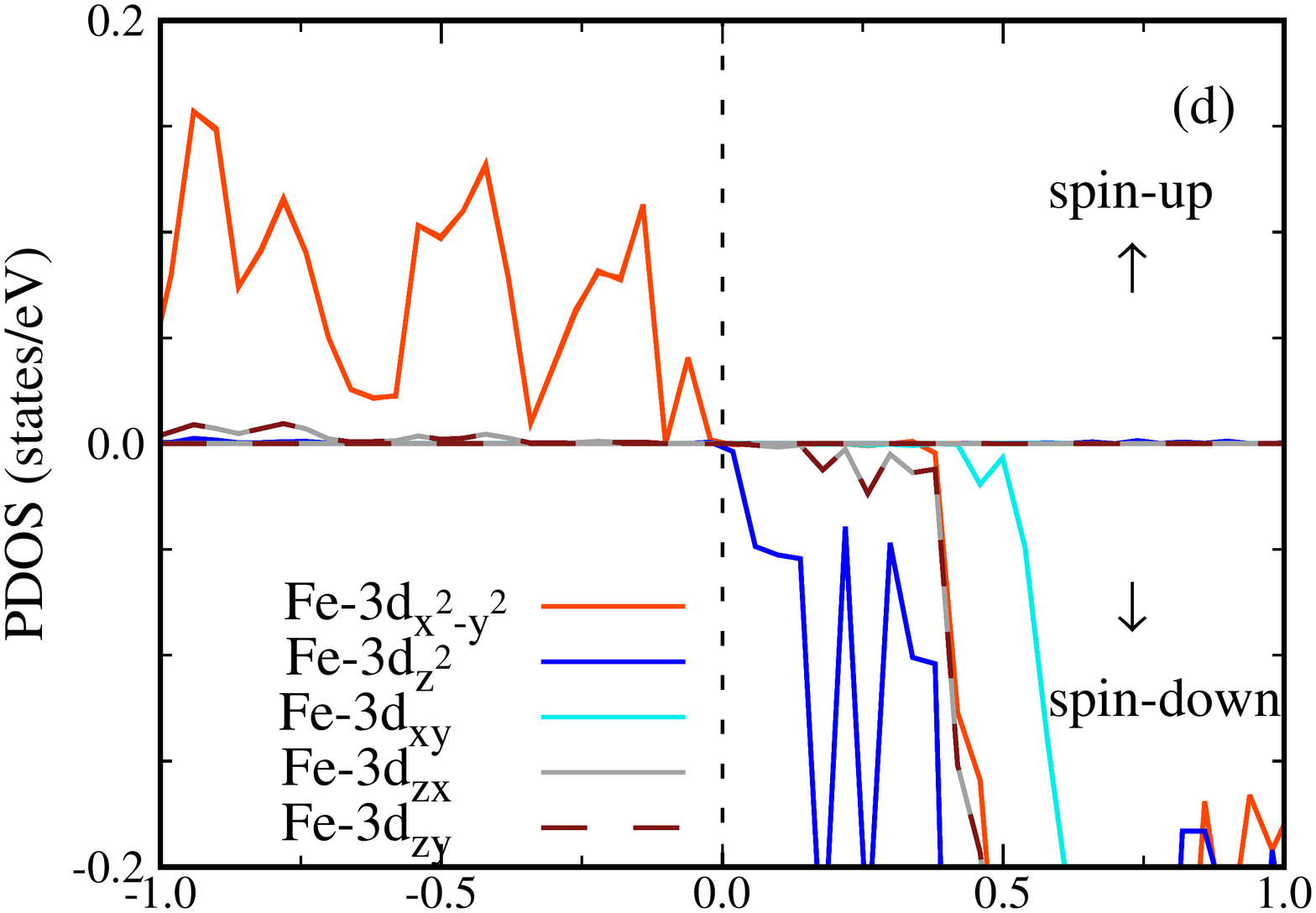}\hspace{-0.5cm}
	\includegraphics[width=0.32\textwidth,height=0.30\textwidth]{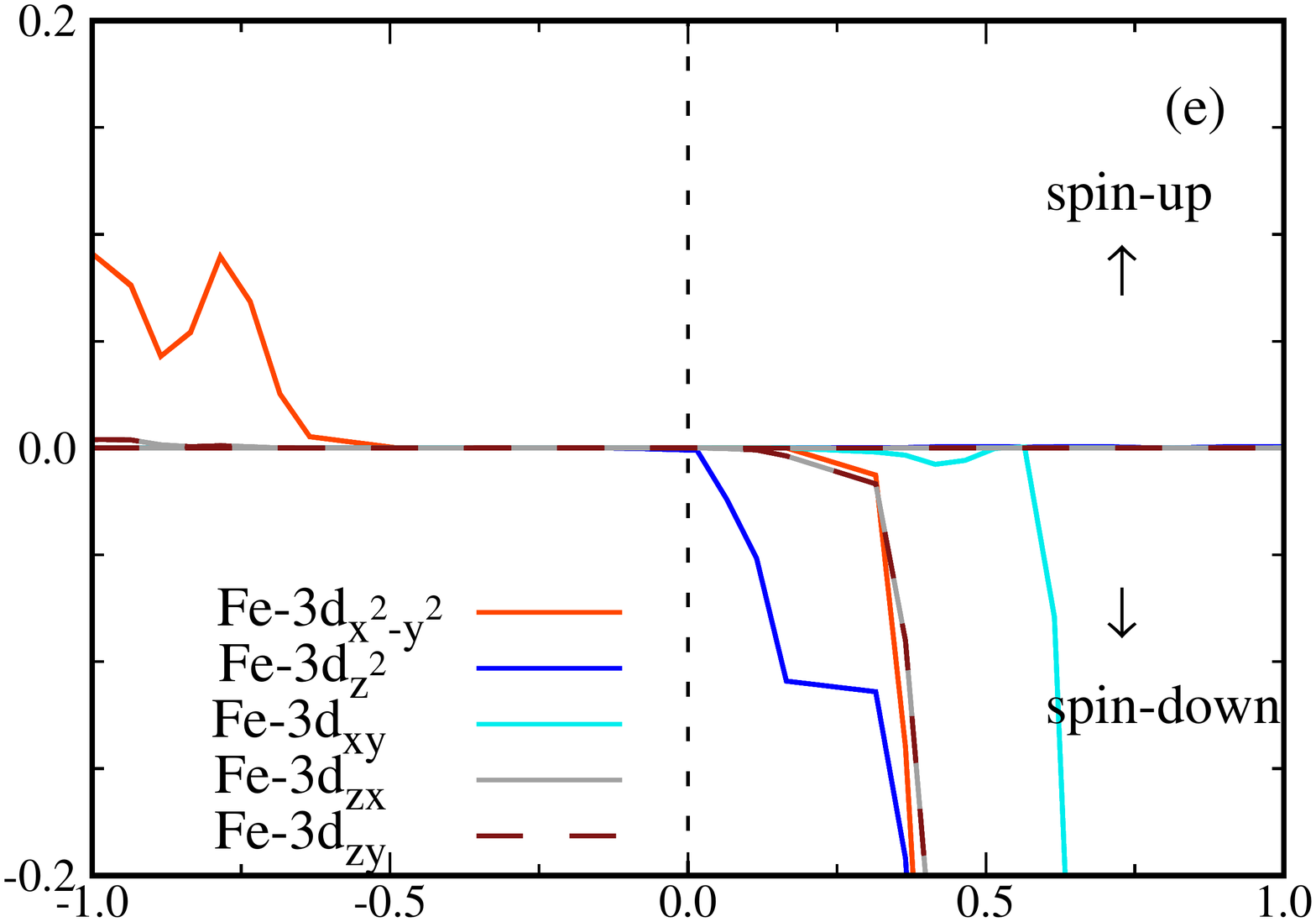}\hspace{-0.5cm}
	\includegraphics[width=0.32\textwidth,height=0.30\textwidth]{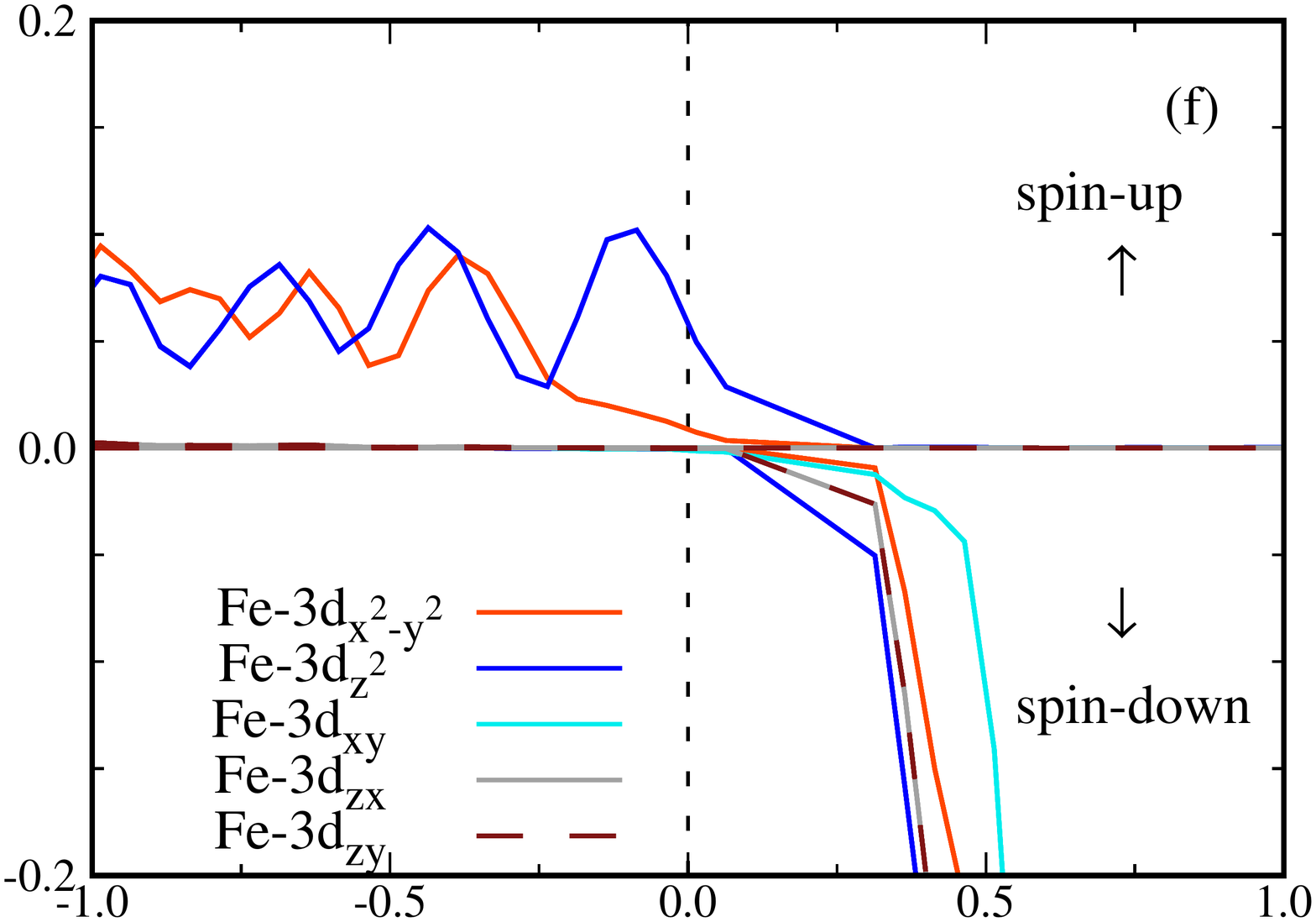}\vspace{-0.5cm}\\
	\includegraphics[width=0.32\textwidth,height=0.30\textwidth]{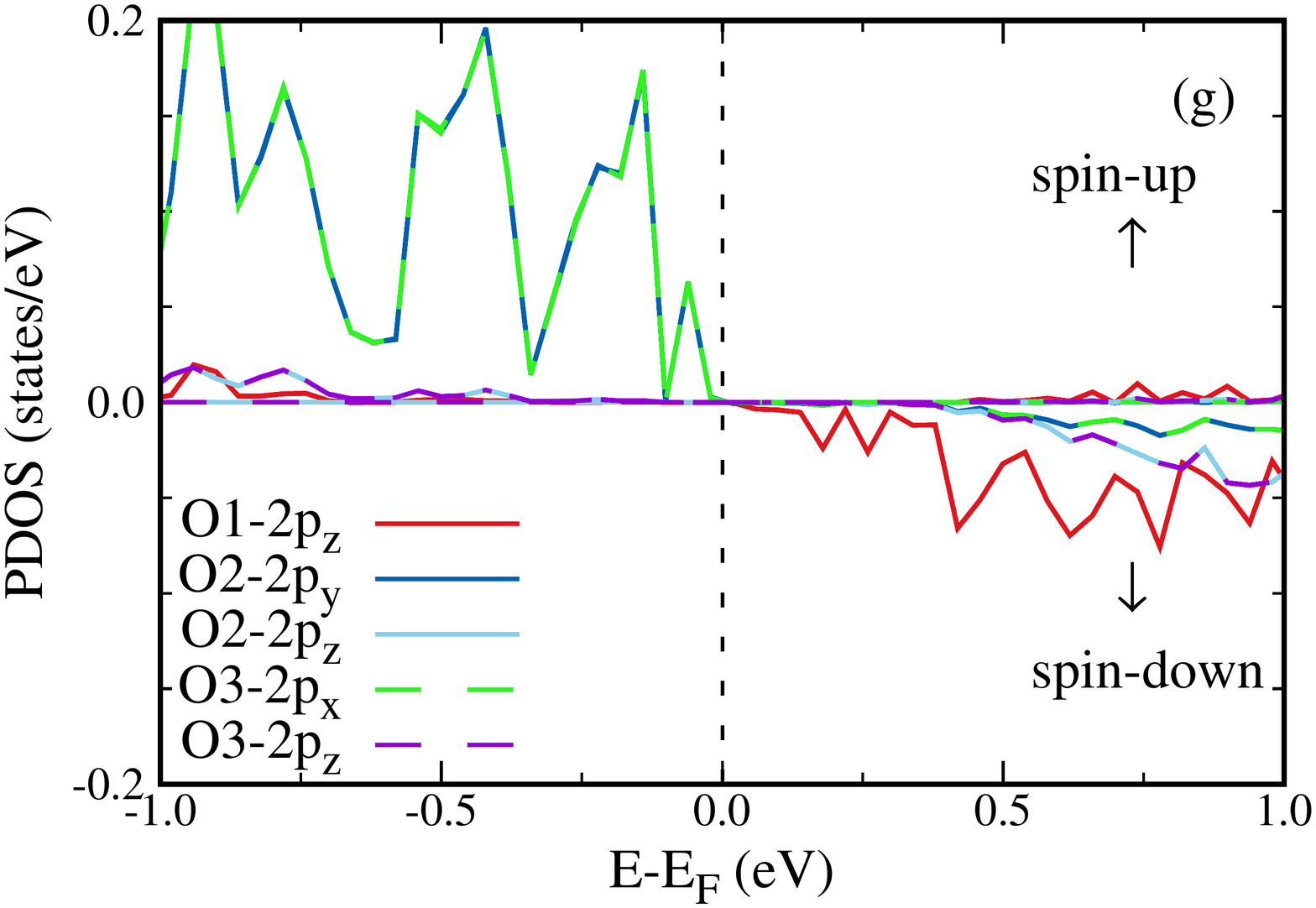}\hspace{-0.5cm}
	\includegraphics[width=0.32\textwidth,height=0.30\textwidth]{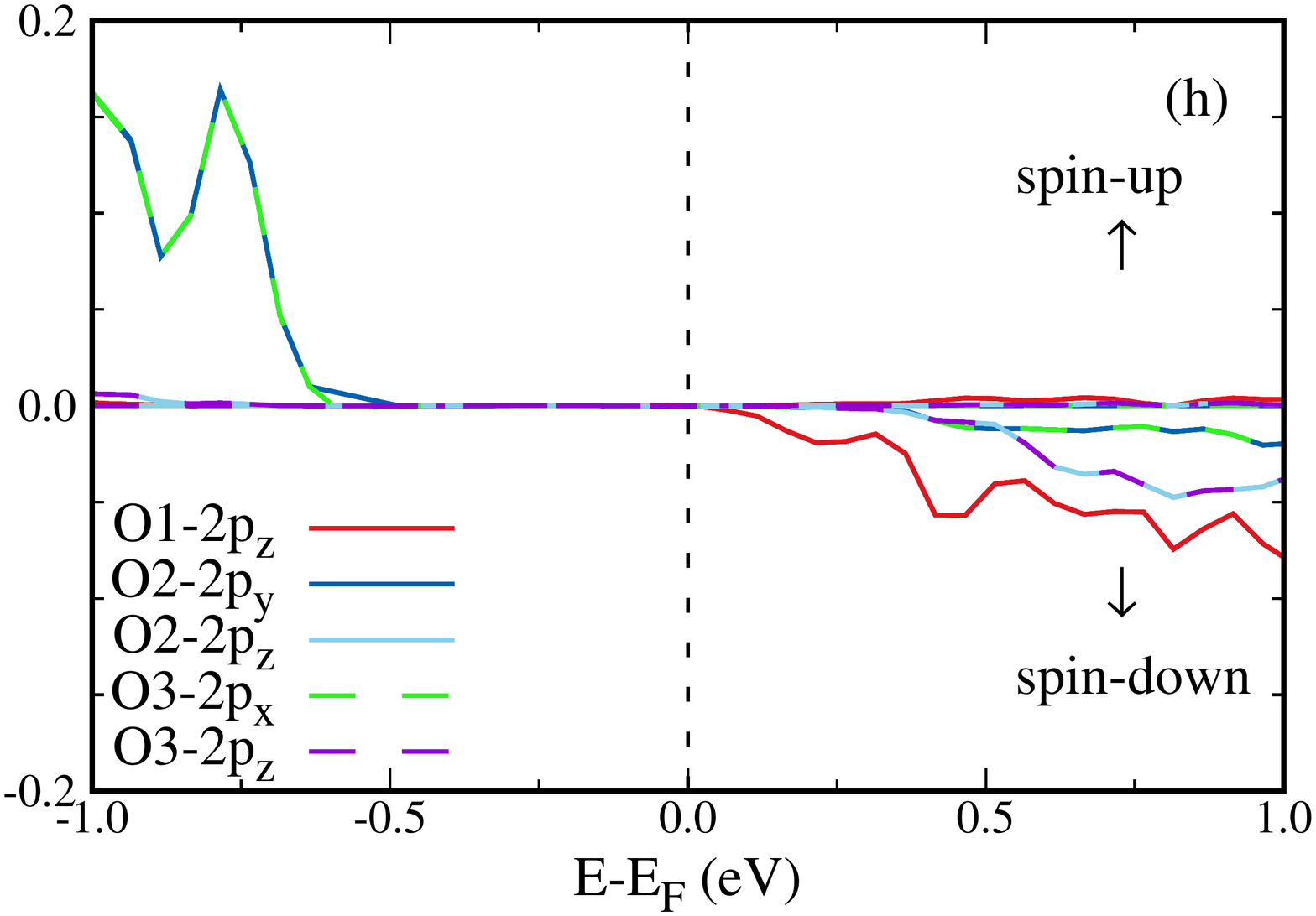}\hspace{-0.5cm}
	\includegraphics[width=0.32\textwidth,height=0.30\textwidth]{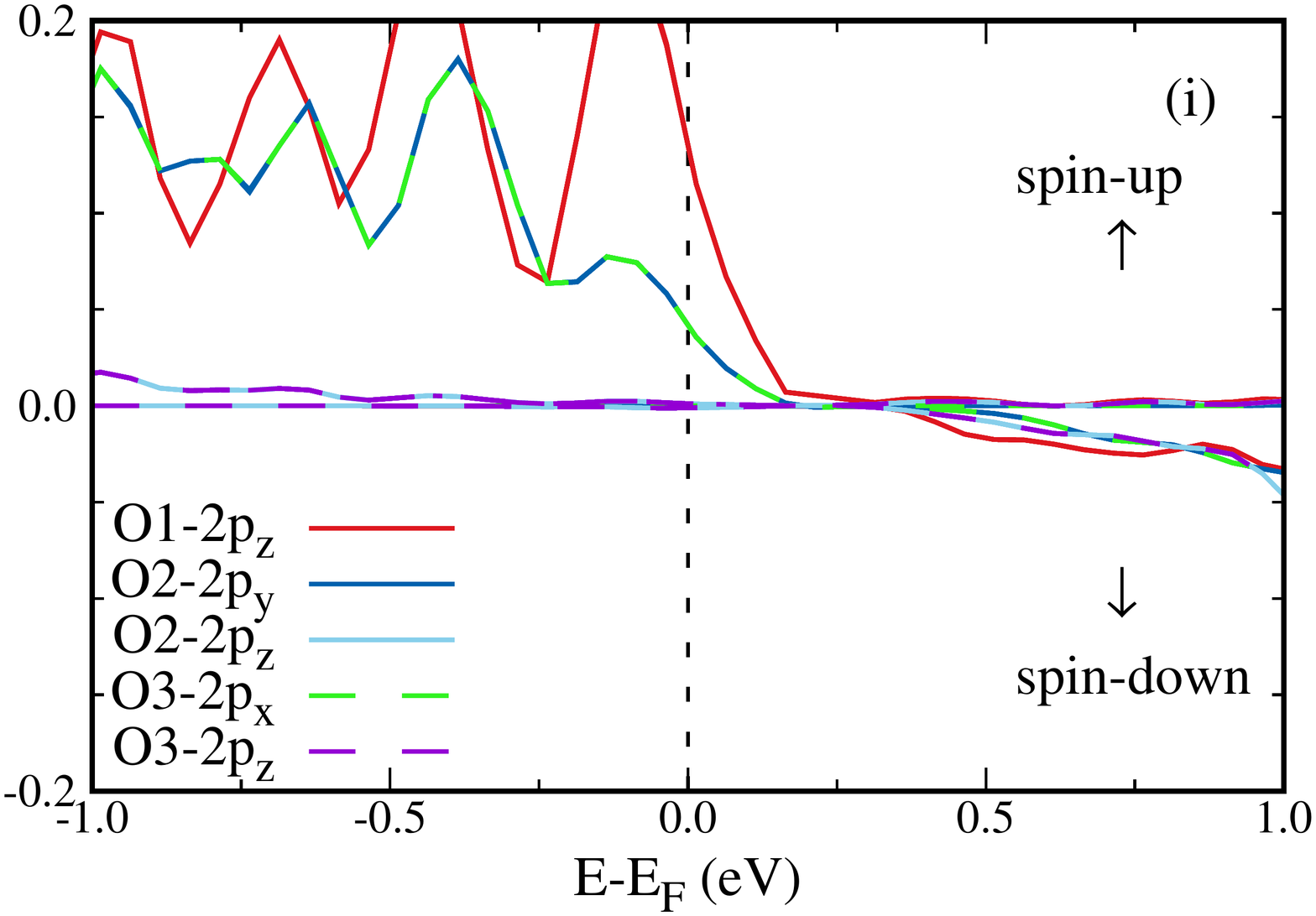}
\vspace{-0.7cm}	
\caption{The top panel \big((a), (b), (c)\big) corresponds to the FM phase in Bi-$6p_x,6p_y,6p_z$. 
The middle panel \big((d), (e), (f)\big)
corresponds to Fe-$3d-{e_g},t_{2g}$ and the lower panel \big((g), (h), (i)\big) corresponds to 
O1, O2, O3-$2p_x,2p_y,2p_z$.}
\label{Fig:fm-pdos}
\end{figure*}
Since our primary interest is in the nature of the FM phase, in the following sections we present the results 
of T-BFO with only FM ordering. To understand the relation between the energy of a given structure and its $c/a$ ratio, 
we have studied the Fe-$\textrm{O}_{e}$-Fe angle, and the O1-Fe and Fe-$\textrm{O}_{e}$ bond length of each structure 
after performing 
the ionic relaxation. Here  $\textrm{O}_{e}$ represent the equatorial oxygen atoms. These results are presented in 
Table~\ref{Table:Fe-O-Fe-bond}. The direct correlation between the energy and the Fe-$\textrm{O}_{e}$-Fe angle is 
quite apparent. Similar to the energy, from structure-I to structure-II, Fe-$\textrm{O}_{e}$-Fe angle initially decreases and 
then it increases monotonically. 
The bond length between `Fe' and the two type of oxygen, i.e., the axial oxygen O1 and 
the two equatorial oxygen $\textrm{O}_{e}$ have also been measured. The O1-Fe bond length decreases 
slightly from the structure-I 
to the structure-II and then increases in structure-III. This bond length remains almost unchanged in structure-IV, compared 
to structure-III. The trend is almost identical with the pattern followed by the energy of each structure, that is, 
similar to the energy,
this bond length is the lowest in the case of structure-II. Interestingly, the structure-II shows the half-metallic character.
However, a similar correlation is not observed between the Fe-$\textrm{O}_{e}$ bond length and the energy. 
While the Fe-O2 and Fe-O3 bond lengths are observed to be the same,  the lowest Fe-$\textrm{O}_{e}$ bond length is found 
in the case of 
structure-I. This bond length increases in structure-II and then remains more or less constant till structure-IV.
In the later sections, we are going to explore in detail the relation between the $c/a$ ratio of each structure 
and its electronic properties.
\subsection{Electronic properties}\label{result-discussion:electronic properties}
For a systematic understanding of the electronic properties, we have computed the band structure, 
total density of states (TDOS),
local density of states (LDOS) and the projected density of states (PDOS) for each structure. In Fig.~\ref{Fig:fm-bs}, we have 
presented the spin-resolved band structure of structure-I, II and IV. The fundamental nature of the band structure of 
structure-III is identical with the structure-IV, and it has been studied in Appendix~\ref{Appendix:str3-bs}. 
From the Fig.~\ref{Fig:fm-bs}(a), it is clear that for the highest $c/a$ 
ratio $1.264$, i.e., in structure-I, no states are available at the Fermi energy from either of the spin channel.
However, the band gap is not the same for the two spin channels. The band gap corresponding to the spin-up channel
is estimated to be $\simeq 0.53~\rm{eV}$, while it is $\simeq 1.61~\rm{eV}$ for the spin-down channel. The 
band gaps have been calculated at the zone center. This shows that structure-I is a magnetic semiconductor similar 
to the R3c phase of BFO \cite{rbfo}. From Fig.~\ref{Fig:fm-bs}(c), it is obvious that structure-IV is a metal 
as both the spin channels contribute states at the Fermi level. A similar behavior is 
observed in structure-III (Appendix~\ref{Appendix:str3-bs}). 
\begin{figure*}[htbp!]
	\includegraphics[width=0.24\textwidth,height=0.26\textwidth,trim=1.2cm 0cm 0.1cm 0.9cm,clip=true]{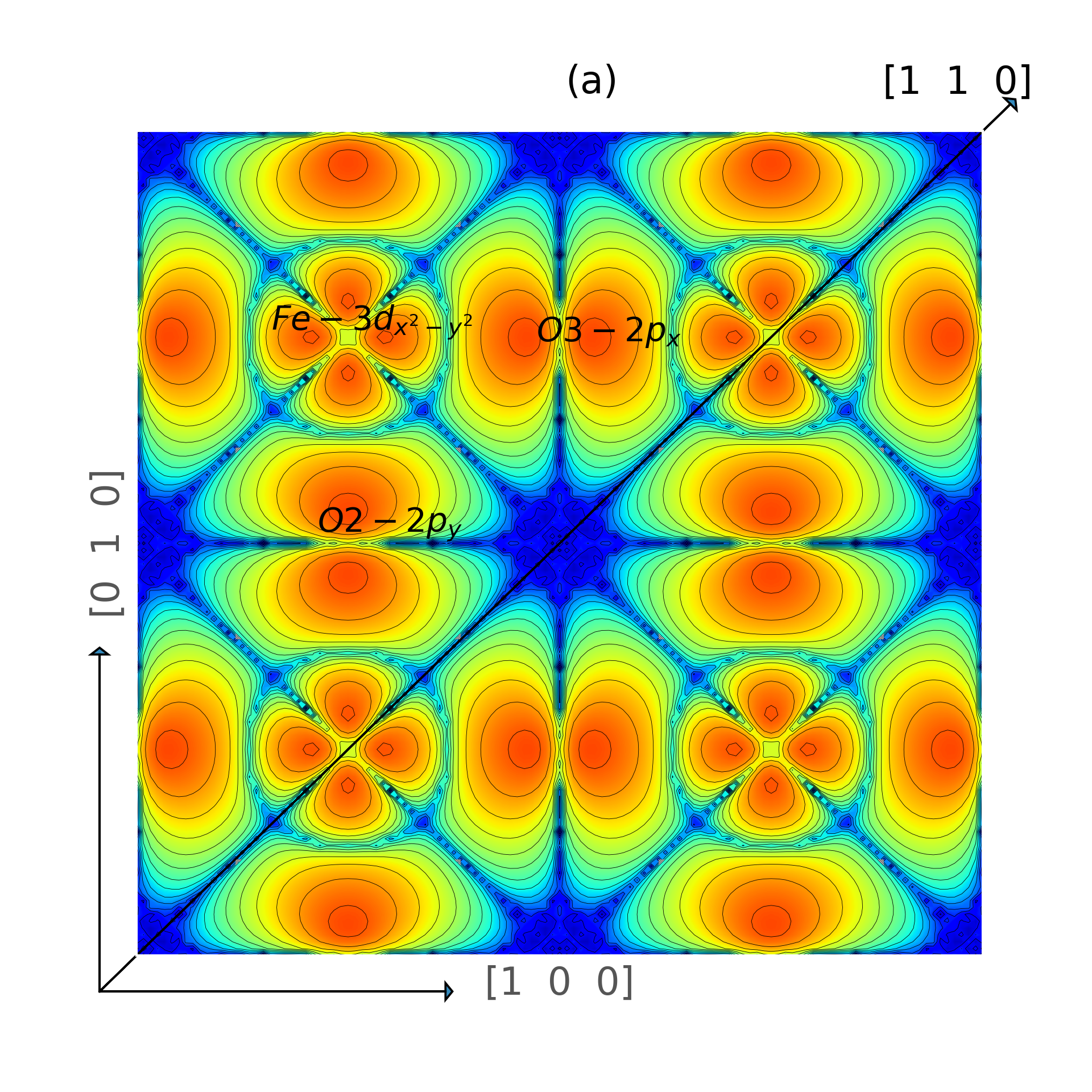}\hspace{-0.33cm}
	\includegraphics[width=0.28\textwidth,height=0.26\textwidth,trim=1cm 0cm 1.5cm 0.9cm,clip=true]{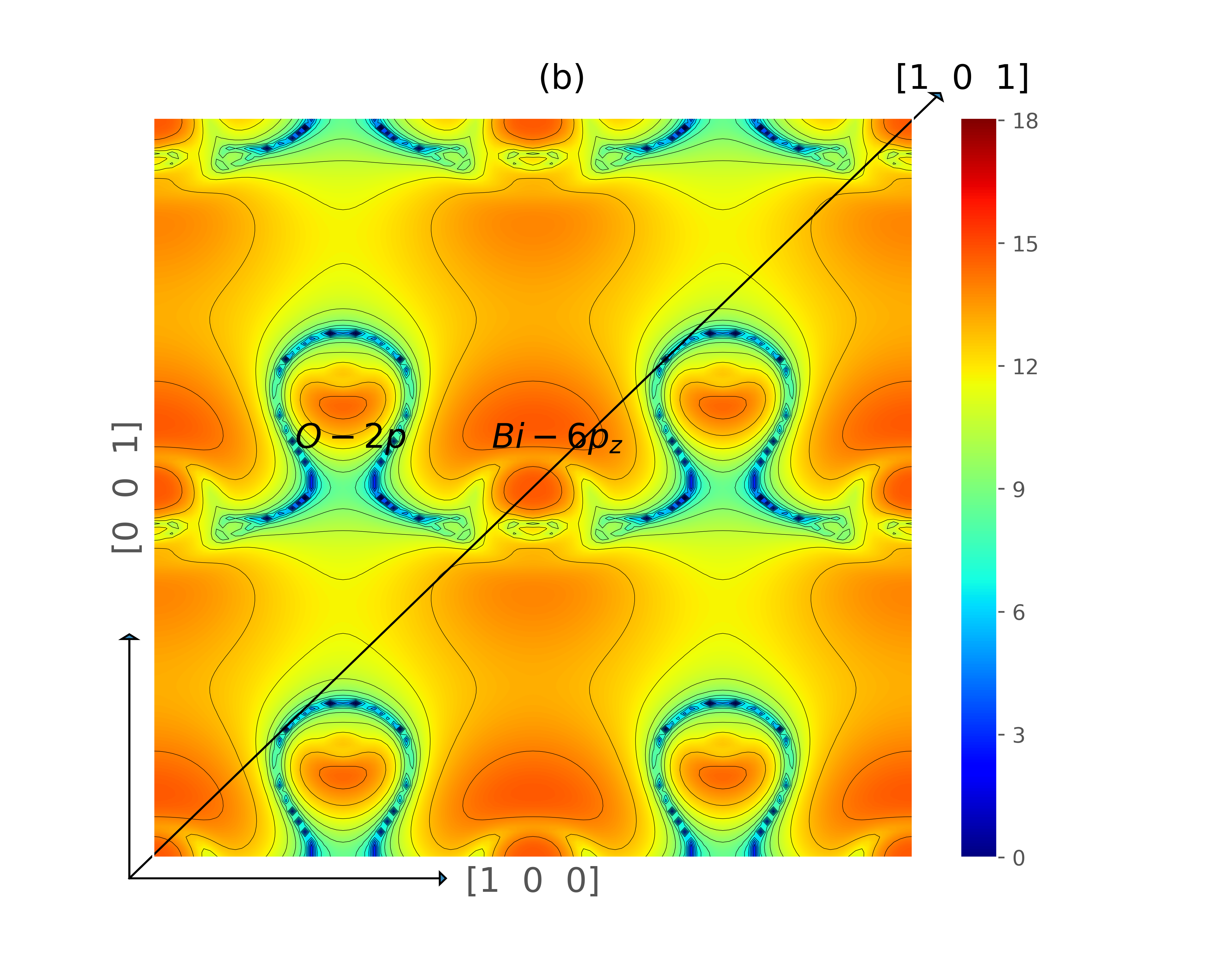}\hspace{-0.4cm}
	\includegraphics[width=0.24\textwidth,height=0.26\textwidth,trim=1.2cm 0cm 0.1cm 0.9cm,clip=true]{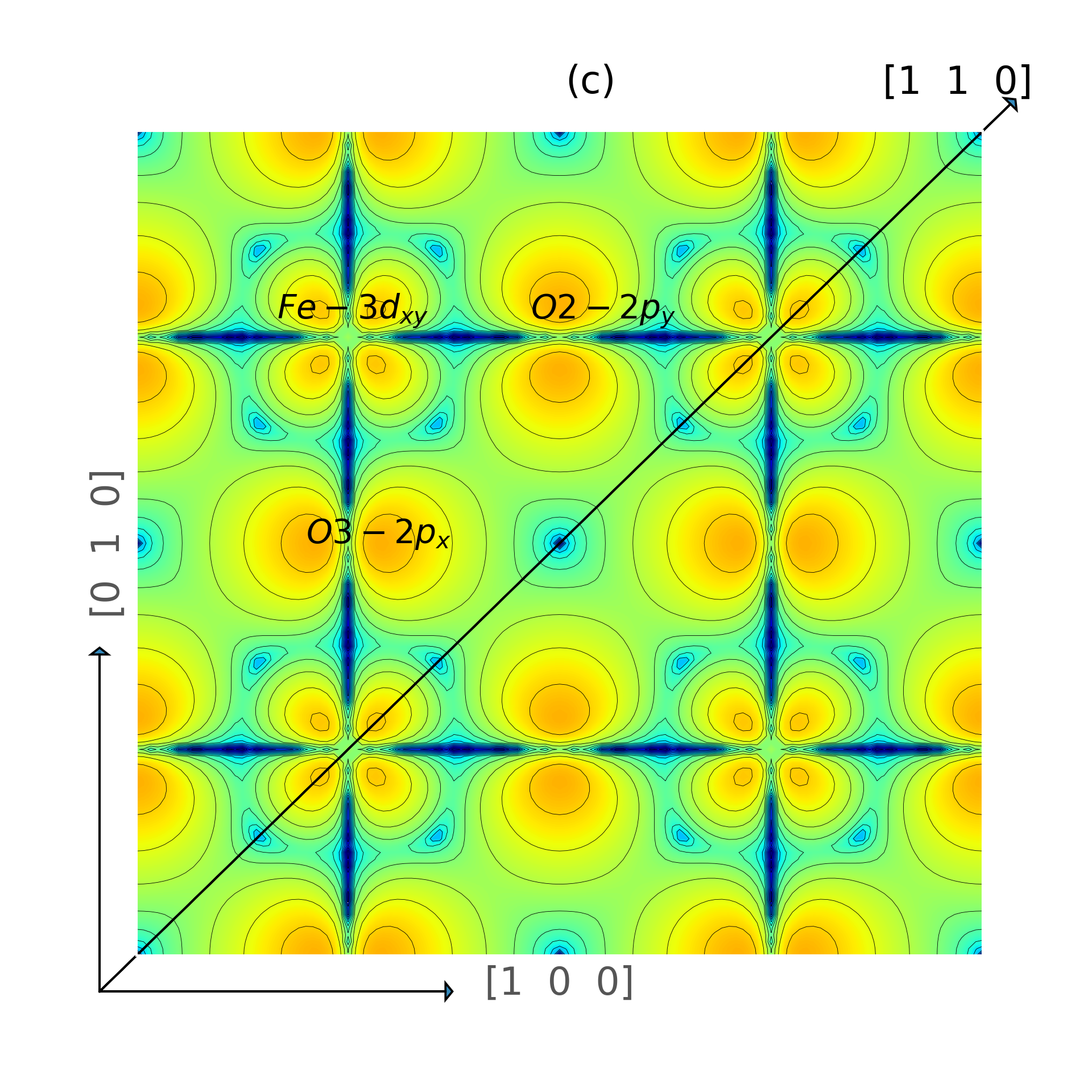}\hspace{-0.33cm}
	\includegraphics[width=0.28\textwidth,height=0.26\textwidth,trim=1cm 0cm 1.5cm 0.9cm,clip=true]{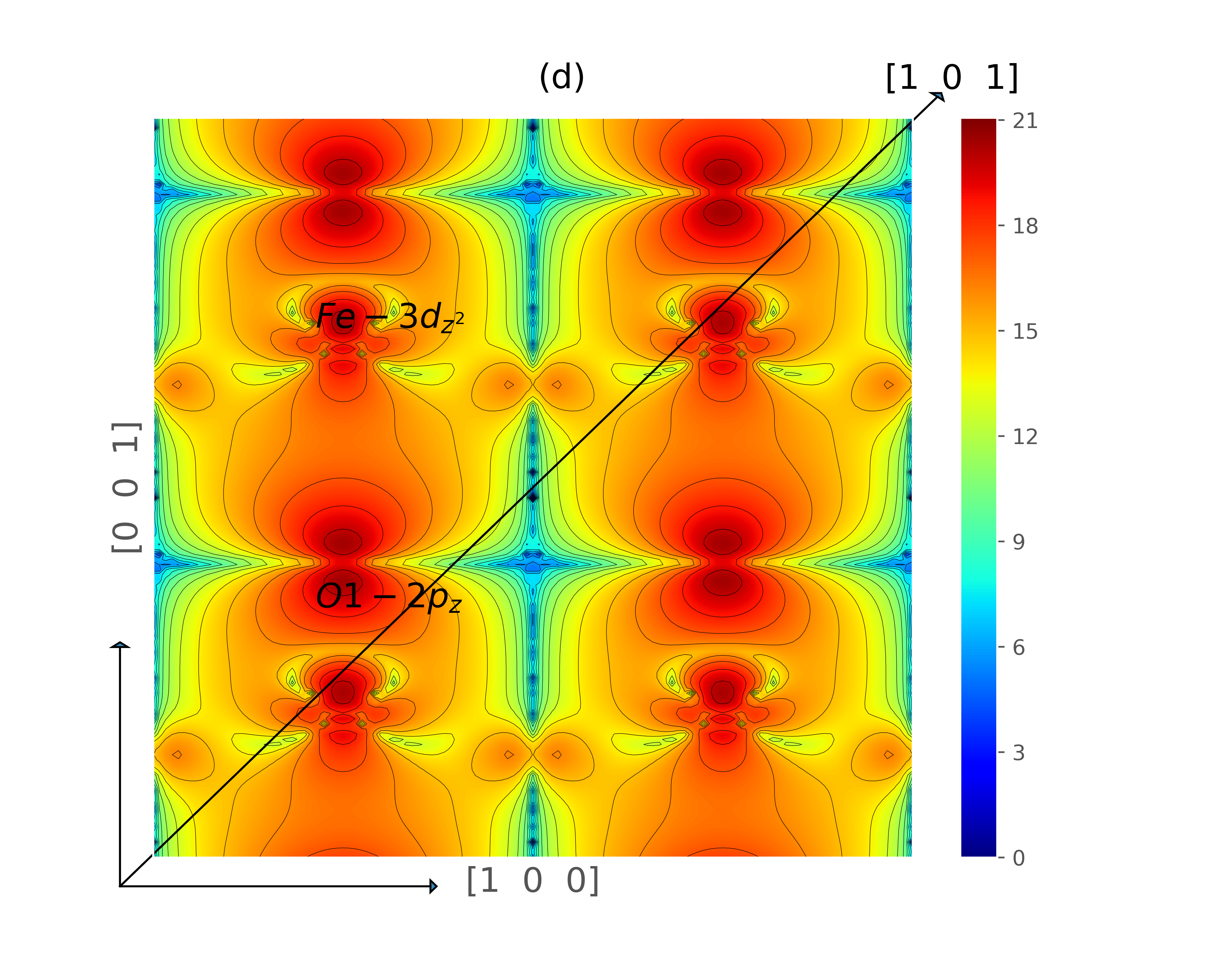}
\vspace{-0.6cm}
\caption{HOMO and LUMO of structure-I with (a) and (b) spin-up, and 
(c) and (d) spin-down respectively.}	
\label{Fig:hl-st1}
\end{figure*}
\begin{figure*}[htbp!]
	\includegraphics[width=0.27\textwidth,height=0.28\textwidth,trim=1.2cm 0.2cm 0.7cm 0.9cm,clip=true]{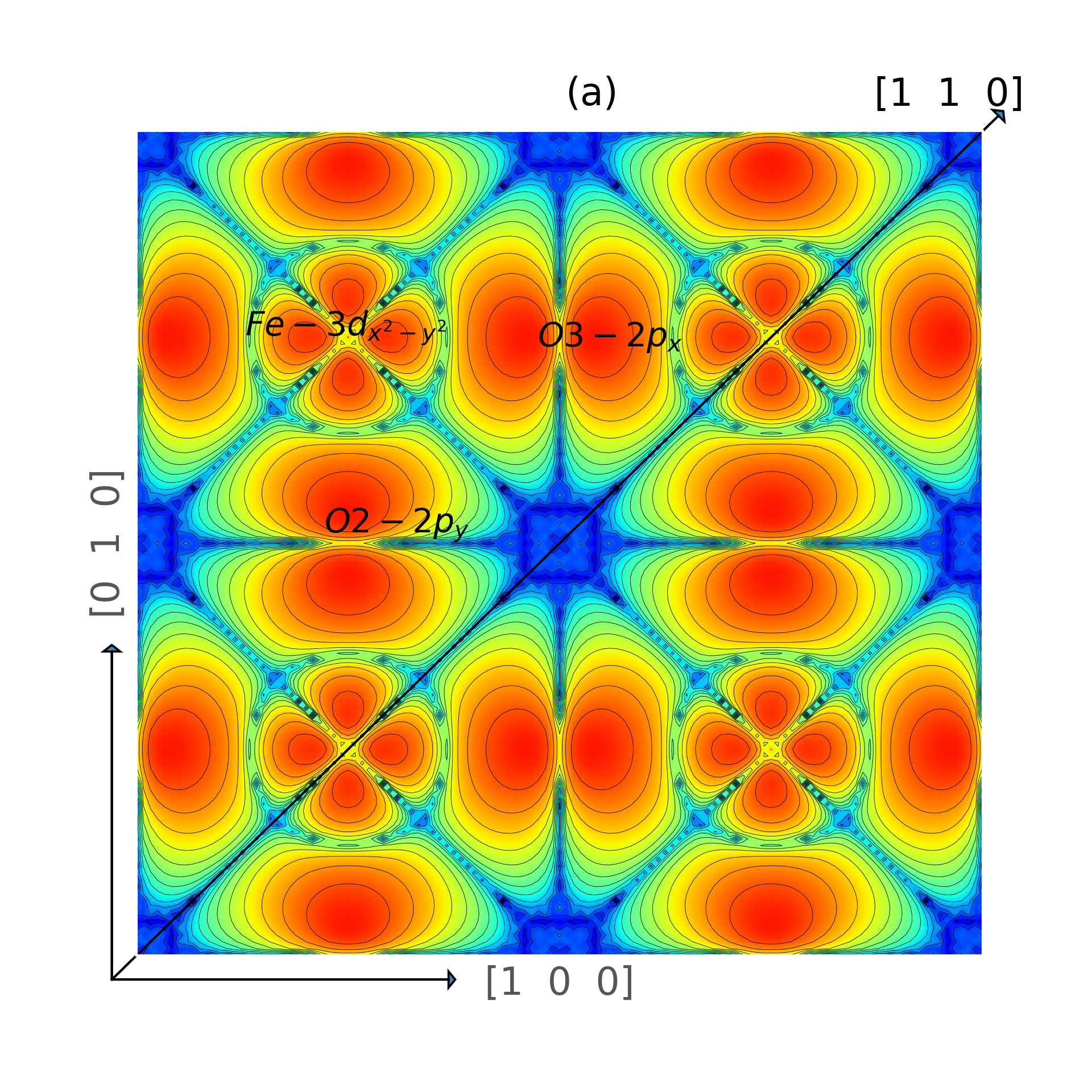}\hspace{-0.15cm}
	\includegraphics[width=0.30\textwidth,height=0.28\textwidth,trim=1.2cm 0.2cm 1.2cm 0.9cm,clip=true]{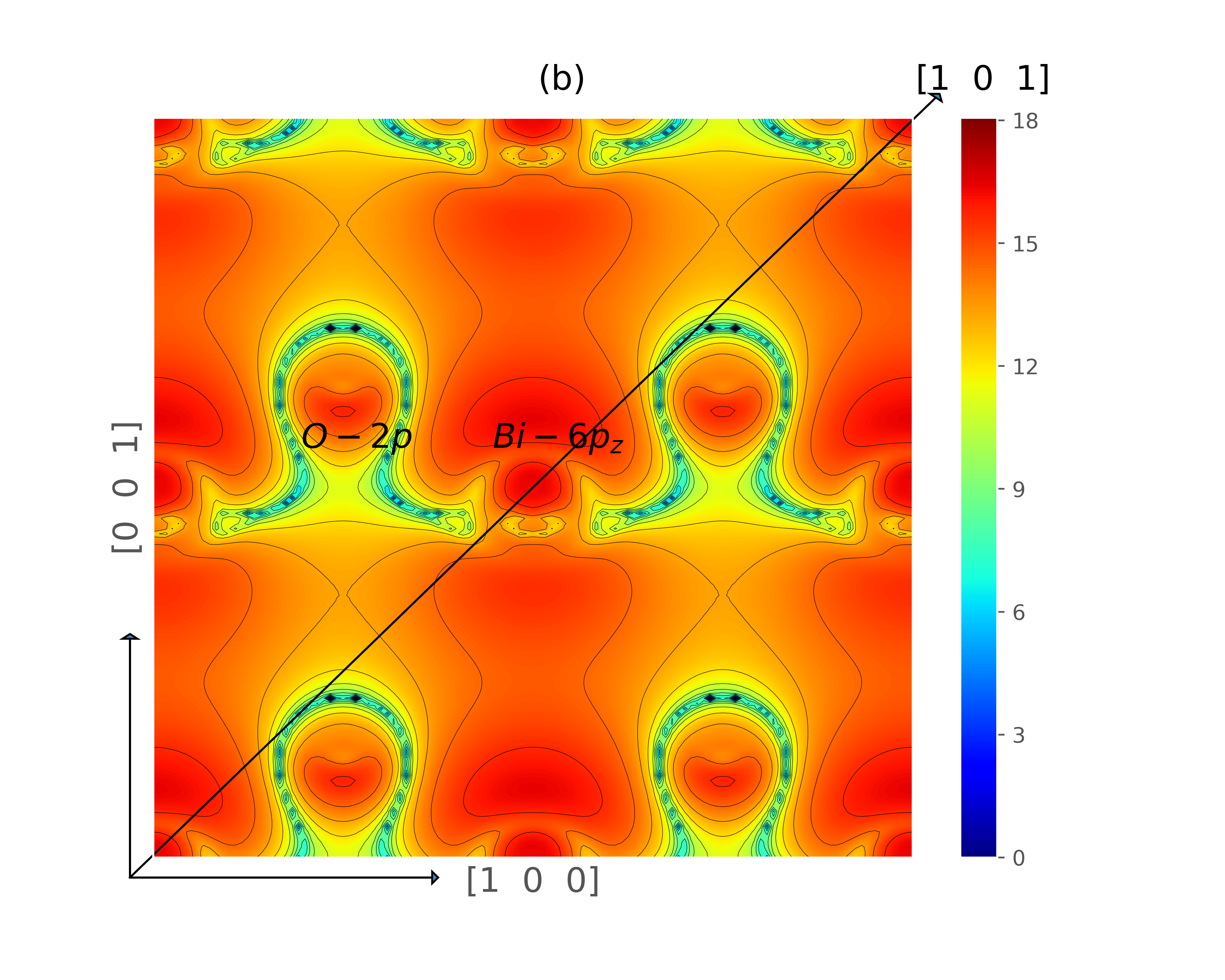}\hspace{-0.15cm}
	\includegraphics[width=0.30\textwidth,height=0.28\textwidth,trim=1.5cm 0.2cm 0.8cm 0.9cm,clip=true]{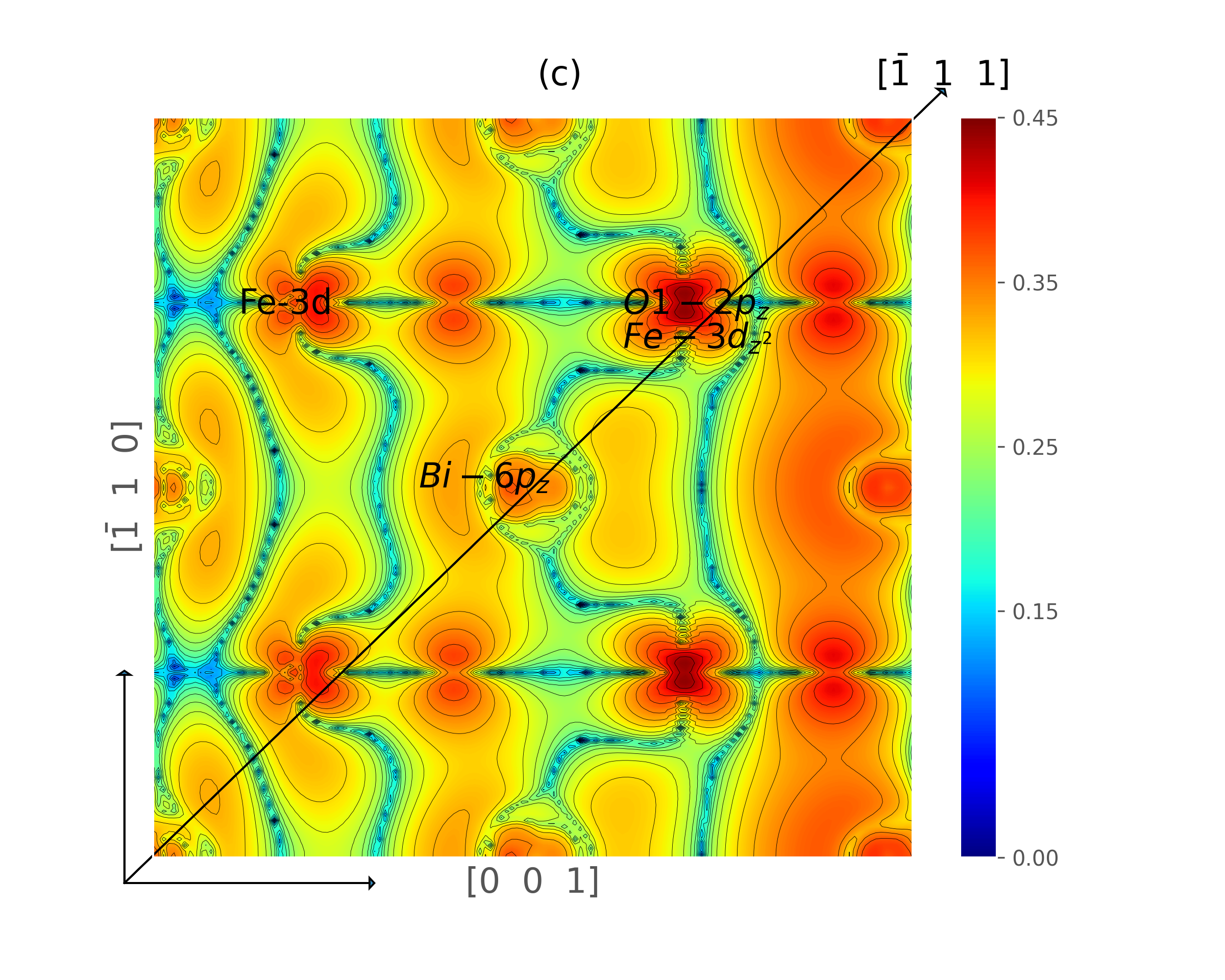}
\vspace{-0.6cm}
\caption{(a) and (b): HOMO and LUMO of structure-I with spin-up, and (c) Molecular orbital
for spin-down.}	
\label{Fig:hl-st2}
\end{figure*}

However, in between these two limits of the $c/a$ ratio, in the intermediate structure with $c/a=1.233$, i.e.,
the structure-II surprisingly has a non-zero contribution at the Fermi energy only from the spin down channel. This makes  
the structure-II behave like a \textit{half-metal}. In this case a band-gap of $\simeq 0.89~\rm{eV}$ is observed in the 
spin-up channel. The band structure is presented in Fig.~\ref{Fig:fm-bs}(b). A closer look at the systematic change 
in the electronic properties with respect to the volume of each structure reveals a fascinating phenomena, which has 
not been reported so far in a pure compound to the  best of our knowledge. A simple 
analysis shows that the volume of the structure-II is the highest, while the structure-I has a lower volume than the 
structure-II. Structure-III and IV occupy almost the same volume and is less than that of the structure-I.
This indicates that the change from the semiconducting state of the structure-I to the half-metallic phase of 
the structure-II is 
quite counter-intuitive, since the volume effectively increases from the structure-I to the structure-II. 
It is generally expected 
that under pressure (with reduction of volume), the insulator/semiconductor turns into a metal. Interestingly, our results 
indicate an opposite trend to this general expectation. Furthermore, the semiconducting state of the structure-II becomes 
a metal in the structure-III and remains a metal in structure-IV. This change is in tune with the general expectation of the
insulator/semiconductor to metal transition under pressure, as the volume of structure-III/IV is lower than that 
of the structure-II.
Interestingly, this kind of metallic to insulating/semiconducting and back to metallic transition has been 
observed in pure Li, which becomes a semiconductor at $80~\rm{GPa}$ \cite{Shimizu} and then it becomes a metal once 
again at $120~\rm{GPa}$ \cite{Matsuoka}. Experimental evidences of this kind of counter-intuitive metal to insulator transition under 
high pressure have also been reported earlier in Na \cite{Ma-Sodium}, FCC-Ca \cite{Dunn-FCC-Ca}, \cite{Stager-FCC-Ca}, and 
Ni \cite{McMahan-Ni}, and in the binary compound $\rm{CLi}_4$ \cite{Jin-Chen-CLi4}.

The total density of states (TDOS) corresponding to the above mentioned 
band structures have been shown in Fig.~\ref{Fig:fm-tdos}. In Fig.~\ref{Fig:fm-tdos}(a), 
the TDOS has been presented for the structure-I, and it is clear that a band gap exists for both the 
spin components, although they are unequal. It is also clear that the band gap is smaller for the spin-up component. 
The TDOS of structure-II has been presented in Fig.~\ref{Fig:fm-tdos}(b). While there are no states at the Fermi energy 
from the spin-up channel, there exists a small but finite number of states from the spin-down component, which is the typical 
signature of a half-metal. It is clear from the Fig.~\ref{Fig:fm-tdos}(c), that both the spin components contribute to the 
conductivity in the case of structure-IV, although the contribution is more from the spin-up component. 

In order to examine the orbital contribution of different atoms to the electronic properties in the T-BFO 
structures, we have computed the local density-of-states (LDOS). These results are presented in Fig.~\ref{Fig:fm-ldos}. 
In Fig.~\ref{Fig:fm-ldos}(a), the results for the structure-I has been presented. It is clear that there are no states at the 
Fermi energy and there exists a band gap for both the spin channels. However, as the $c/a$ ratio is lowered from structure-I
to structure-II, one can observe in Fig.~\ref{Fig:fm-ldos}(b) that a finite amount of electronic states appear at the Fermi 
energy. Interestingly, only the spin-down channel of the Bi-$6p$ orbital appears to make the contribution to the TDOS
 at the Fermi energy. In the molecular orbital analysis \textcolor{red}{(Sec.~\ref{MO-CD-H})},
we have found contributions from the Fe-$3d$ and O-$2p$ orbitals.   
In Fig.~\ref{Fig:fm-ldos}(c), we have presented the results for the structure-IV, which is found to be a metal from its
band structure. However, there are some interesting features to observe in the charge carriers and its spin type.  
`Bi', more specifically Bi-$6p$ orbital, contributes electron carriers from both the 
spin channels, while Fe-$3d$ and O-$2p$ orbitals contribute hole carriers to the system only from the spin-up channel.  

It is important to note that, like the ferromagnetic rhombohedral phase\cite{rbfo}, in Fig.~\ref{Fig:fm-ldos}, 
we can also observe an indication of strong hybridization between the Fe-$3d$ and O-$2p$ orbitals in both the valence and the 
conduction bands. Furthermore, a hybridization can also be observed between Bi-$6p$ and O-$2p$ in the conduction band. The 
hybridization between Fe-$3d$ and O-$2p$ is mainly found in the spin-up channel of the valence band and in the spin-down 
channel of the conduction band close to the Fermi energy. It is also important to note that the Bi-$6s$ orbital contribution 
is found far below the Fermi level at around $-1.5eV$, which indicates that these Bi-$6s$ lone pair electrons are localized 
in this system. However, Bi-$6s$ orbital takes part in the hybridization process. In Sec.~\ref{MO-CD-H}, 
we have discussed the type 
of hybridization and the role of specific atomic orbitals in more details.\\
\begin{figure*}[htbp!]
	\includegraphics[width=1.0\textwidth,height=0.30\textwidth,trim=2.9cm 1cm 1.5cm 0.5cm,clip=true]{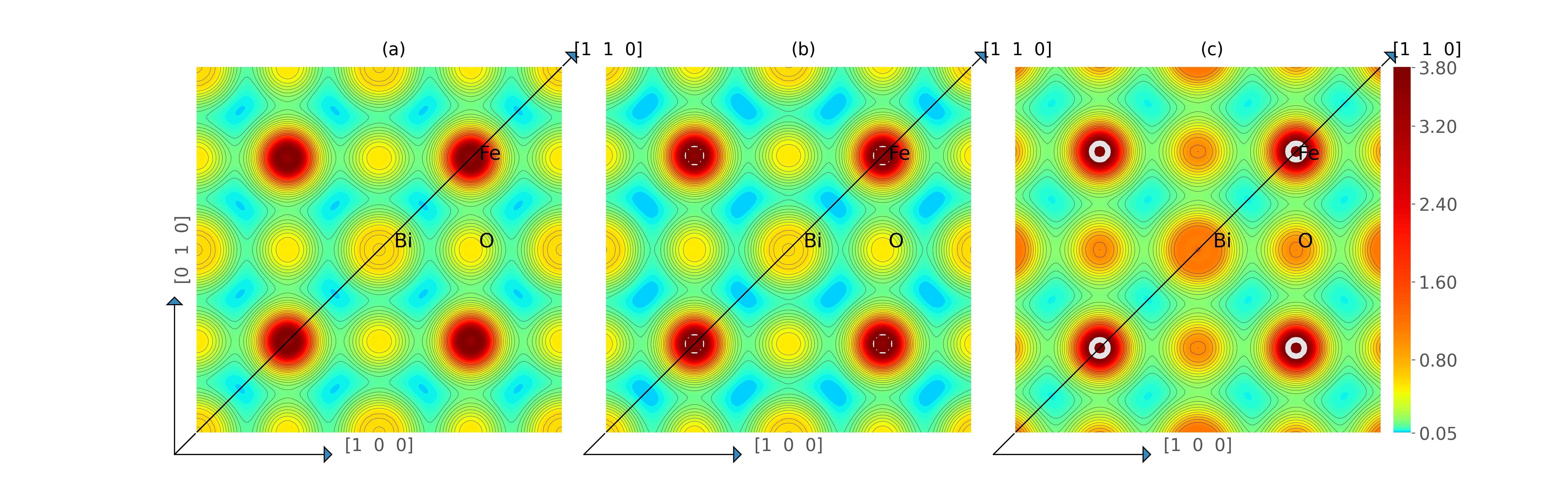}
\vspace{-0.7cm}
\caption{The charge density (CD) distributions at $\{001\}$-plane for (a) structure-I, (b) structure-II, and 
(c) structure-IV respectively.}	
\label{Fig:fm-cd}
\end{figure*}
To further understand the contribution of the individual orbitals, we have calculated the projected density of 
states (PDOS), and the results have been presented in Fig.~\ref{Fig:fm-pdos}. 
From Figs.~\ref{Fig:fm-pdos}(a),(d), and (g), it is clear 
that there is a band gap in both the spin channels. It is also clear that in this structure the band just
above the Fermi energy has 
contributions from Bi-$6p_z$, Fe-$3d_{z^2}$ and O1-$2p_z$ orbitals. Interestingly, all of these orbitals belong 
to the spin-down channel. On the other hand, the band just 
below the Fermi energy has contributions only from Fe-$3d_{x^2-y^2}$, O2-$2p_y$, and O3-$2p_x$. 
These orbitals belong to the spin-up channel. It is interesting to note there is 
no contribution from the Bi atom in this band. Surprisingly, as the $c/a$ ratio is lowered in the structure-II, 
the spin-down channel 
of Bi-$6p_z$ moves below the Fermi level. This is very evident from Fig.~\ref{Fig:fm-pdos}(b). The spin-down channels of the 
Fe-$3d_{z^2}$ and O1-$2p_z$ orbitals also move below the Fermi energy. However, they have vanishingly small but finite 
contributions 
to the PDOS. Their presence has been detected by the molecular orbital study, which has been discussed in Sec.~\ref{MO-CD-H}. 
Interestingly, the spin-up channels of the orbitals, which were the closest to the Fermi energy in the structure-I 
are pushed further 
down in energy in the structure-II, giving rise to the \emph{half-metallic} character to the T-BFO. As 
the $c/a$ ratio is lowered 
further in the structure-IV, both the spin channels of the Bi-$6p_z$ orbital move further below the Fermi energy, 
while only the 
spin-up component of the $e_g$ band of Fe along with the  O1-$2p_z$, O2-$2p_y$, and O3-$2p_x$ orbitals 
crosses over the Fermi level, 
thereby making the structure-IV a ferromagnetic metal. Furthermore, it is also clear from the Figs.~\ref{Fig:fm-pdos}(c), (f) 
and (i), that `Bi' contributes only electron carriers of both types of spin to the system, while the other orbitals 
mentioned above contribute 
only hole carriers, that too from the spin-up channel, to the system. It is also important to note that the Bi-$6p_z$ 
orbital consistently 
moves towards the Fermi level from above and then crosses over the Fermi energy as the $c/a$ ratio is decreased gradually. 
On the other hand, 
other orbitals systematically moves closer to the Fermi energy from below and finally crosses over with the decrease 
in the volume from the 
structure-I to the structure-IV.    
\begin{table}[htbp!]
\caption{Estimation of the total magnetic moment (TM) and the individual magnetic moments (in $\mu_b$)
contributed by `Bi', `Fe' and `O' corresponding to the G-type anti-ferromagnetic (G-AFM) phase.}
\begin{center}
\begin{tabular}{c c c c c c c }
\hline 
Structures  & str-I  & str-II  & str-III & str-IV   \\
c/a ratio   & 1.264  & 1.233   & 1.049   & 1.016  \\
\hline
TM          & 0.0000 & 0.0000  &  0.0000 &  0.0000 \\
Bi          & $\mp0.0000$  &$\mp0.0000$  & $\mp0.0000$ & $\mp0.0000$  \\
Fe          & $\mp3.5996$  & $\mp3.6031$ & $\mp3.6152$ & $\mp3.6127 $ \\
O           & $\mp0.1823$  & $\mp0.2195$ & $\mp0.1417$ &  $\mp0.1343$ \\
\hline
\label{Table:G-AFM-contribution}
\end{tabular}
\end{center}
\end{table}
\begin{table}[htbp!]
\caption{Estimation of the total magnetic moment (TM) and the individual magnetic moments (in $\mu_b$)
contributed by `Bi', `Fe' and `O' corresponding to the 
ferromagnetic (FM) phase.}
\begin{center}
\begin{tabular}{c c c c c c c }
\hline
Structures  & str-I  & str-II & str-III & str-IV   \\

c/a ratio   & 1.264  & 1.233  & 1.049 & 1.016 \\
\hline
TM          & 40.000 &  40.000 & 39.950 & 39.830 \\
Bi          & 0.0146 & 0.0117  & 0.0071 & 0.0063  \\
Fe          & 3.7300 & 3.6900  & 3.8300 & 3.8200  \\
O           & 0.2240 & 0.2584  & 0.2453 & 0.2450 \\
\hline
\label{Table:FM-contribution}
\end{tabular}
\end{center}
\end{table}

\subsection{Magnetic properties}\label{result-discussion:magetism}
In this section, we discuss the magnetic properties of the T-BFO structure and its dependence on the $c/a$ ratio.
In the FM phase, the total magnetic moment (TM) is found to be around $40~\mu_{b}$ for all of the four structures, which 
also agrees well with the results of the ferromagnetic R3c phase of BFO \cite{rbfo}. Contribution of different atoms to the 
magnetization in the FM phase are listed in the Table~\ref{Table:FM-contribution}. To understand 
the crucial differences in the contribution towards magnetization from different atoms compared with the anti-ferromagnetic 
phase, we have presented a similar analysis for the G-AFM phases in the Table~\ref{Table:G-AFM-contribution}.
The major difference that we have found is that in the case of G-AFM phases `Bi' doesn't contribute anything to the 
total magnetic moment for all the four structures, while in the case of FM phases it has a finite contribution to 
the total magnetic moment. This behavior is similar compared 
to the R3c structure \cite{rbfo}. In the T-BFO, the contribution decreases monotonically with the $c/a$ ratio from 
structure-I to structure-IV. As expectedly, Fe's contribution towards the total magnetic moment has been found to be the largest 
among all the atoms in both the FM and G-AFM phases. Furthermore, the Fe's magnetic moment has been found to be larger in the FM 
phase, and this trend is also similar to the rhombohedral phase \cite{rbfo}. However, in the T-BFO phase Fe's magnetic moment has
been observed to be lower compared to the rhombohedral phase.

A similar scenario is found in the case of `O' atom as well, which agrees with the R3c structure, i.e., it contributes
much more towards the total magnetization in the FM phase as compared to the G-AFM phase \cite{rbfo}. However, in comparison to the 
rhombohedral structure, the magnetic moment of the `O' atom is significantly higher in the G-AFM phase of the T-BFO structure, while 
they are comparable in the FM phase. In this phase, in the case of structure-IV and structure-III, the TM is less 
than $40~\mu_{b}$, while in the structure-II it is exactly $40~\mu_{b}$. This is consistent with our conclusions based on the band 
structure and the TDOS results presented in Fig.~\ref{Fig:fm-bs} and Fig.~\ref{Fig:fm-tdos}. It is well known  
\cite{pickett} that in the case of a 
half-metallic phase, the magnetic moment should be an integer multiple of $\mu_b$. 
Interestingly, the structure-I, which is found to be a magnetic semiconductor, also has a magnetic 
moment of $40~\mu_b$. This numerical value is identical with the strength of the magnetic moment obtained in the 
magnetic semiconducting state of the rhombohedral BFO structure \cite{rbfo}. 
In the case of structure-I, we also would like to point out that the Bi's contribution to the total magnetic moment is 
higher in the magnetic semi-conducting phase of the T-BFO structure compare to the R3c structure \cite{rbfo}.   

\subsection{Molecular orbital, Charge density and Hybridization:}\label{MO-CD-H}
For further understanding about the orbital contributions,
we have investigated the nature of the molecular orbitals in the vicinity of Fermi energy.
In Fig.~\ref{Fig:hl-st1},  we have presented the highest occupied molecular orbital (HOMO) and the 
lowest unoccupied molecular orbital (LUMO) for both the spin channels corresponding to the structure-I.
From Fig.~\ref{Fig:hl-st1}(a) it is clear that the HOMO corresponding to the spin-up channel has contribution from the  
Fe-$3d_{x^2-y^2}$ orbital and the $2p$ orbitals of the equatorial oxygen, more specifically from the O2-$2p_y$  and O3-$2p_x$
orbitals. This is consistent with the PDOS results presented in Figs.~\ref{Fig:fm-pdos}(a), (d), and (g). In the LUMO, i.e., 
in Fig.~\ref{Fig:hl-st1}(b) contributions of Bi-$6p_{z}$ and a tiny contribution from the equatorial oxygens can be seen. 
In the spin-down channel, contributions of the Fe-3$d_{xy}$ orbital and the equatorial oxygens are found in the HOMO, 
shown in Fig.~\ref{Fig:hl-st1}(c), while the Fe-3$d_{z^2}$ orbital and the O1-2$p_z$ contribute to the LUMO, 
plotted in Fig.~\ref{Fig:hl-st1}(d). 
In the case of structure-II, for the spin-up channel, the composition of the HOMO and the LUMO shown in 
Figs.~\ref{Fig:hl-st2}(a) and (b) respectively
is identical with 
the same of the structure-I. In the case of spin-down channel which shows the conducting property, we have plotted 
the molecular orbital which is shown in Fig.~\ref{Fig:hl-st2}(c). In this case, the contributions from Bi-$6p_{z}$, Fe-3$d_{z^2}$ 
and the O1-$2p_z$ are very prominent and retain their own character while  the contributions from equatorial oxygens and other 
Fe-3$d$ orbitals i.e., Fe-3$d_{x^2-y^2}$ and Fe-3$d$-$t_{2g}$ orbitals have lost their own characters due to strong hybridization.

Furthermore, the electronic charge density denotes the nature of the bonding among the different atoms.
The charge density calculations along the $\{001\}$ plane gives the details of the sharing of the chemical/ionic bonding 
between `Bi', `Fe' and `O' ions. From the charge density plots of Fig.~\ref{Fig:fm-cd}, we find that 
in the case of structure-I and structure-II the charges are localized which shows the charge sharing is less
between `Bi', `Fe' and `O'. But in the lowest $c/a$ ratio, i.e., in case of the structure-IV, the charges are 
delocalized, which indicates greater charge sharing between `Bi', `Fe' and `O'. 
This gives the confirmation of the ferromagnetic metallic phase which we obtained earlier in the case 
of structure-IV.
In general, the amount of charge sharing between the ions, i.e., the anion and the cation decide the 
type of bonding that exists in the system whether it is ionic or covalent. In this case, we find greater charge difference 
between `Bi' and `O'. This type of charge difference has also been found between `Fe' and `O'. Hence we can conclude 
that the `Bi' and `Fe' cations create ionic bond with electronegative oxygen. 
The perovskite $\rm{BiFeO_3}$ can be viewed as a structure with a central metal (either $\rm{Bi}$ or $\rm{Fe}$)
surrounded by electron rich ligands, i.e, $\rm{O}$. Here, in this case, we find that `Fe' plays the role of principal
central metal atom as it has more contribution than `Bi' to the hybridization, which we find from L${\ddot{o}}$wdin charges. 
In this study, modified $sp^2d^2$ hybridization 
has been found in all the cases which we have discussed in Appendix~\ref{Appendix:Lowdin-charge}.
In any such inorganic cluster, 
there is a direct electron donation from the ligand to the metal atom. It may happen that the metal atom 
may give back some electrons to the ligand LUMO. In the case of structure-II, the electron bridging between O-Bi-O 
and O-Fe-O makes the hybrid orbitals, which might play an important role in the half-metallicity.
\section{Conclusion:}\label{Sec:conclusion}
In conclusion, in this work we have used DFT based first-principle calculation to explore the evolution of the
ferromagnetic phase of the tetragonal $\text{BiFeO}_\text{3}$ structure as a function of the $c/a$ ratio. We have found that
the ferromagnetic TBFO structure harbors rich set of electronic phases, including the \emph{half-metallic} phase, 
incidentally which is also the most stable electronic phase among all the possible ferromagnetic phases. 
Additionally, we have also found that the electronic phase shows a half-metal to a insulator, and finally to a metal 
transition on compression, which is contrary to the commonly anticipated behavior of electronic phases under pressure. 
We have found that Bi-$6p_z$ orbital plays a very crucial role in determining the degree of metallicity in the 
ferromagnetic TBO structure. The structure with the highest $c/a$ ratio was found to be a magnetic-semiconductor.
With the reduction of the $c/a$ ratio, initially only one spin channel of the Bi-$6p_z$ orbital crosses the Fermi 
energy from above, providing the half-metallic character to BFO, and finally both the spin channels cross the 
Fermi energy on further reduction of the $c/a$ ratio. From our molecular orbital analyses, we have found contribution 
of the Fe-$3d$ and O-$2p$ orbitals towards the half-metallicity in this system. Since, typically fabricating stable 
structures with half-metallicity experimentally is a challenging task, our findings are going to be quite 
useful, especially from the point of view of spintronic applications, as the tetragonal BFO structures are 
fabricated quite routinely. 
\appendix
\section{Energy vs c/a ratio}\label{Appendix:c/a-vs-ene}
The relative energy is calculated with respect to the G-AFM energy possessed by the structure-II,
\begin{equation}
 \textrm{Relative  Energy} = \frac{\left(E^{structure-x}_{G-AFM/FM} - E^{^{structure-II}}_{G-AFM}\right)}{E_{G-AFM}} 
 \times 100 \nonumber
\end{equation}

where, $E^{structure-x}_{G-AFM/FM}$ is the energy of any structure with either G-AFM or FM ordering. \\

\begin{figure}[htbp!]
	\includegraphics[width=0.50\textwidth,height=0.35\textwidth]{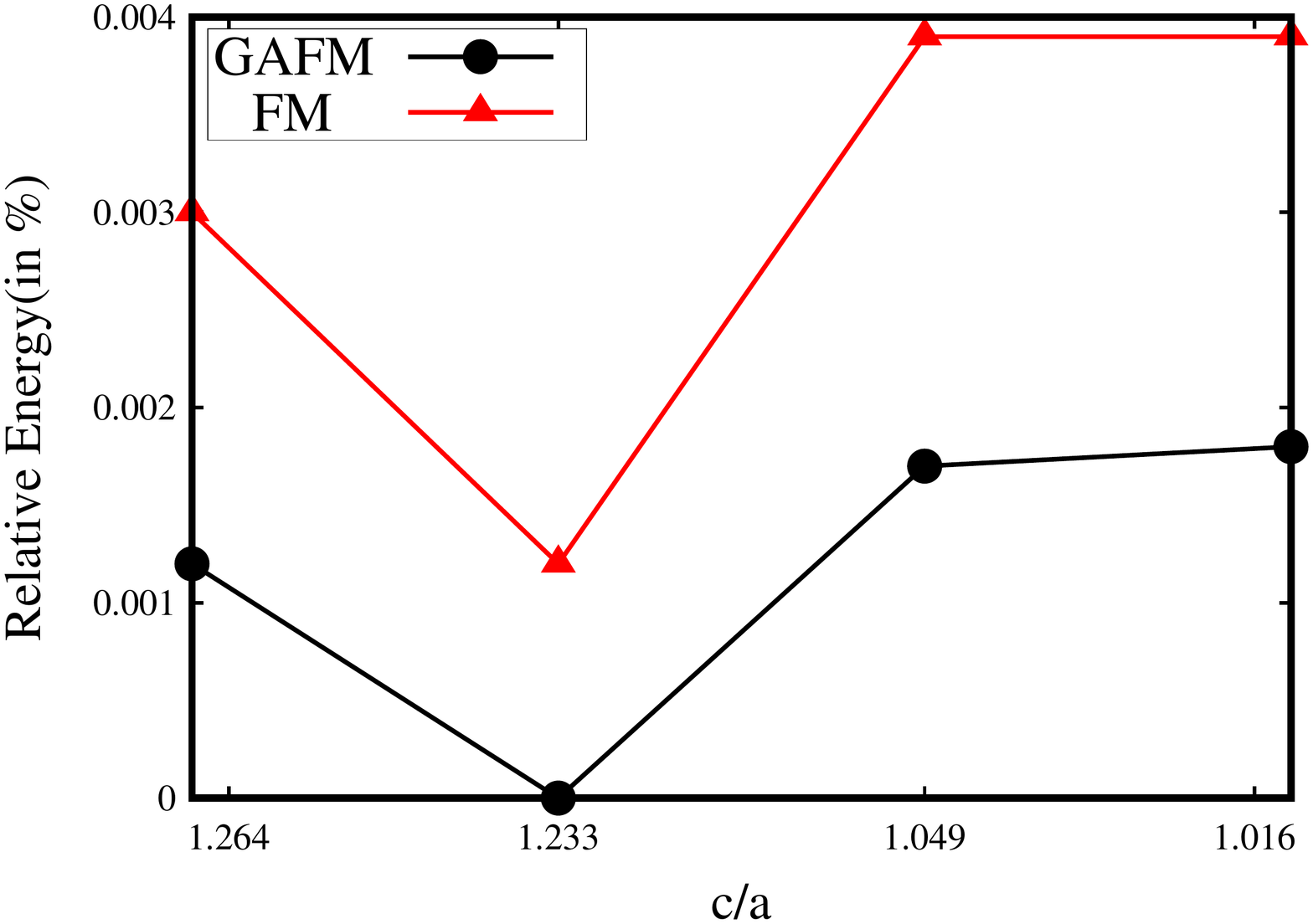}
	\vspace{-0.8cm}
\caption{c/a ratio versus relative energy(in \%), where the relative energy 
of a structure is measured with respect to the most stable G-AFM state, 
that is structure II.}
\label{Fig:ene-vs-c/a}
\end{figure}

The energy of each structure has been represented relative to the energetically most favorable structure, 
that is the structure-II with G-AFM ordering. It is clear from Fig.~\ref{Fig:ene-vs-c/a}, that the
G-AFM type magnetically ordered phase is lower in energy than the corresponding FM phase irrespective of the $c/a$ ratio.
Interestingly, the energy of the FM phase also 
follows a similar pattern to that of the G-AFM phase, that is, the structure-II is the most stable one among all the 
possible structures with FM ordering.  
The lowest energies found in the structure-II with the G-AFM ordering and FM ordering are $-10883.8911~Ry$ and 
$-10883.7638~Ry$ respectively. The energy difference 
for the most stable G-AFM and the FM phase is $0.1273~Ry$ which belongs to the structure-II. The 
 next higher energy differences have been found to be $0.1976~Ry$, $0.2288~Ry$ and $0.2391~Ry$ for the cases of 
 structure-I, structure-IV and structure-III respectively.
\begin{figure*}[htbp!]
 	\includegraphics[width=0.3\textwidth,height=0.25\textwidth,trim=1.5cm -0.5cm 1.5cm 2.0cm,clip=true]{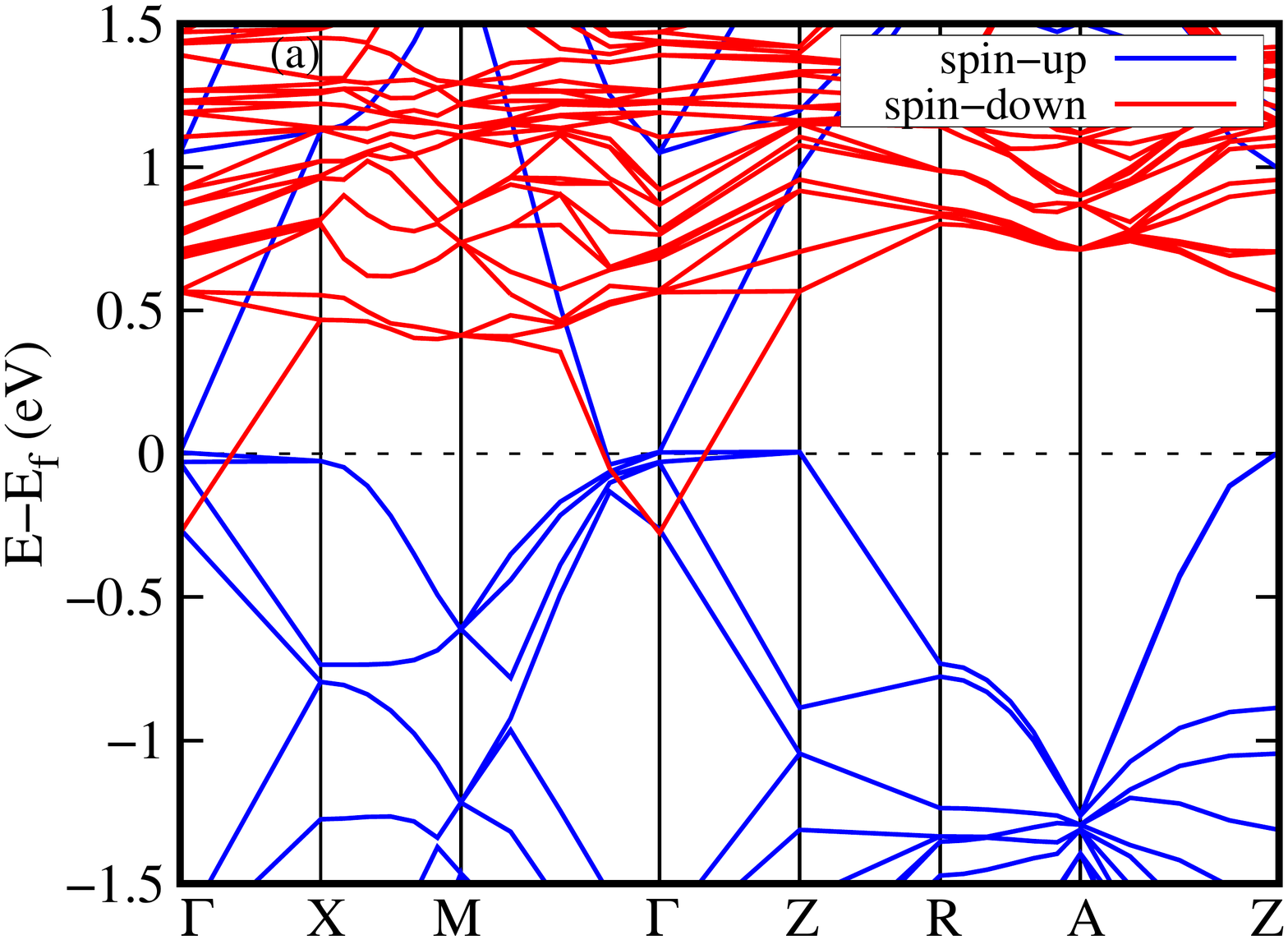}\hspace{-0.1cm}
 	\includegraphics[width=0.3\textwidth,height=0.25\textwidth,trim=1.5cm 1.7cm 1.5cm 2.0cm,clip=true]{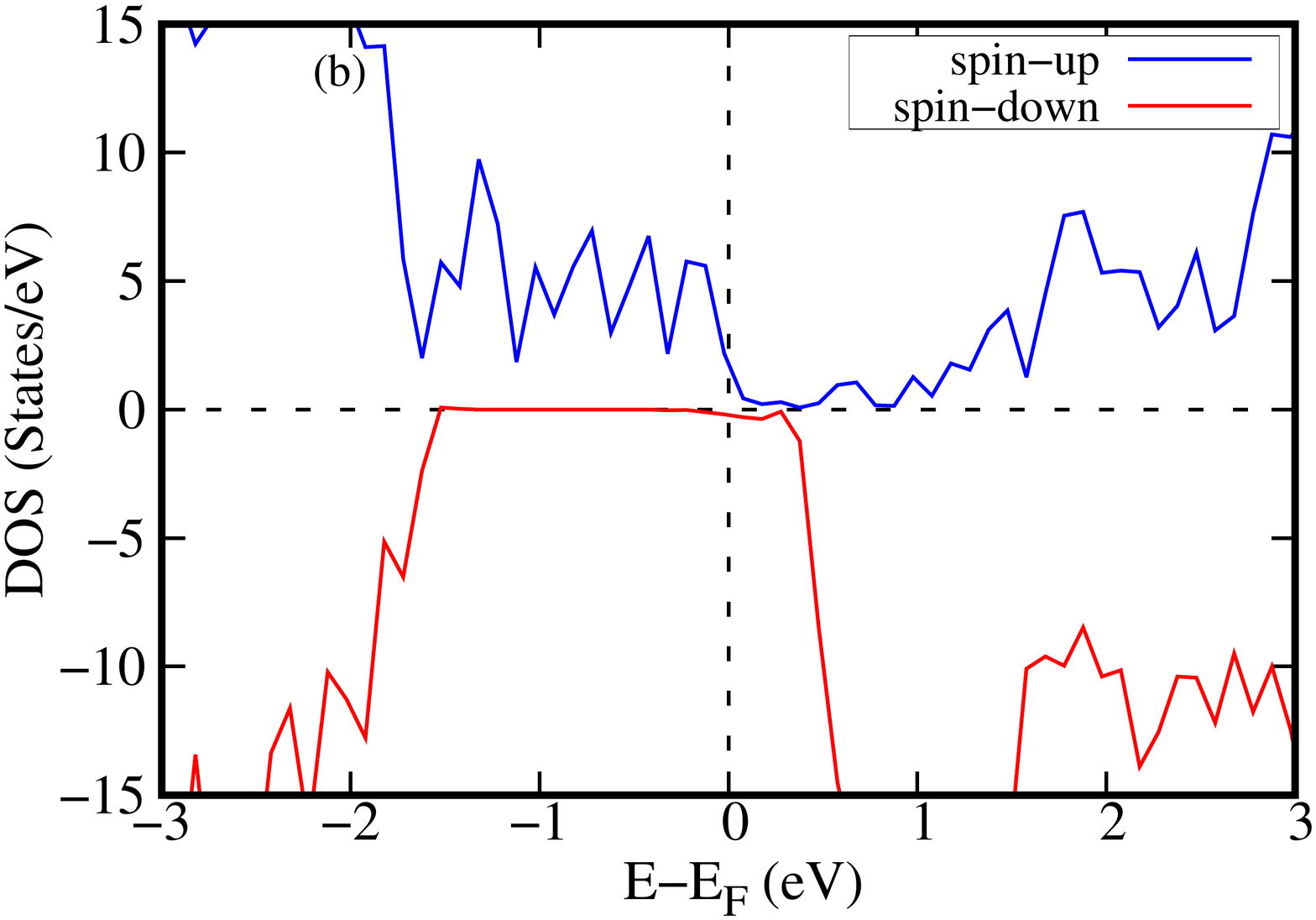}\hspace{-0.1cm}
 	\includegraphics[width=0.3\textwidth,height=0.25\textwidth,trim=1.5cm 1.7cm 1.5cm 2.0cm,clip=true]{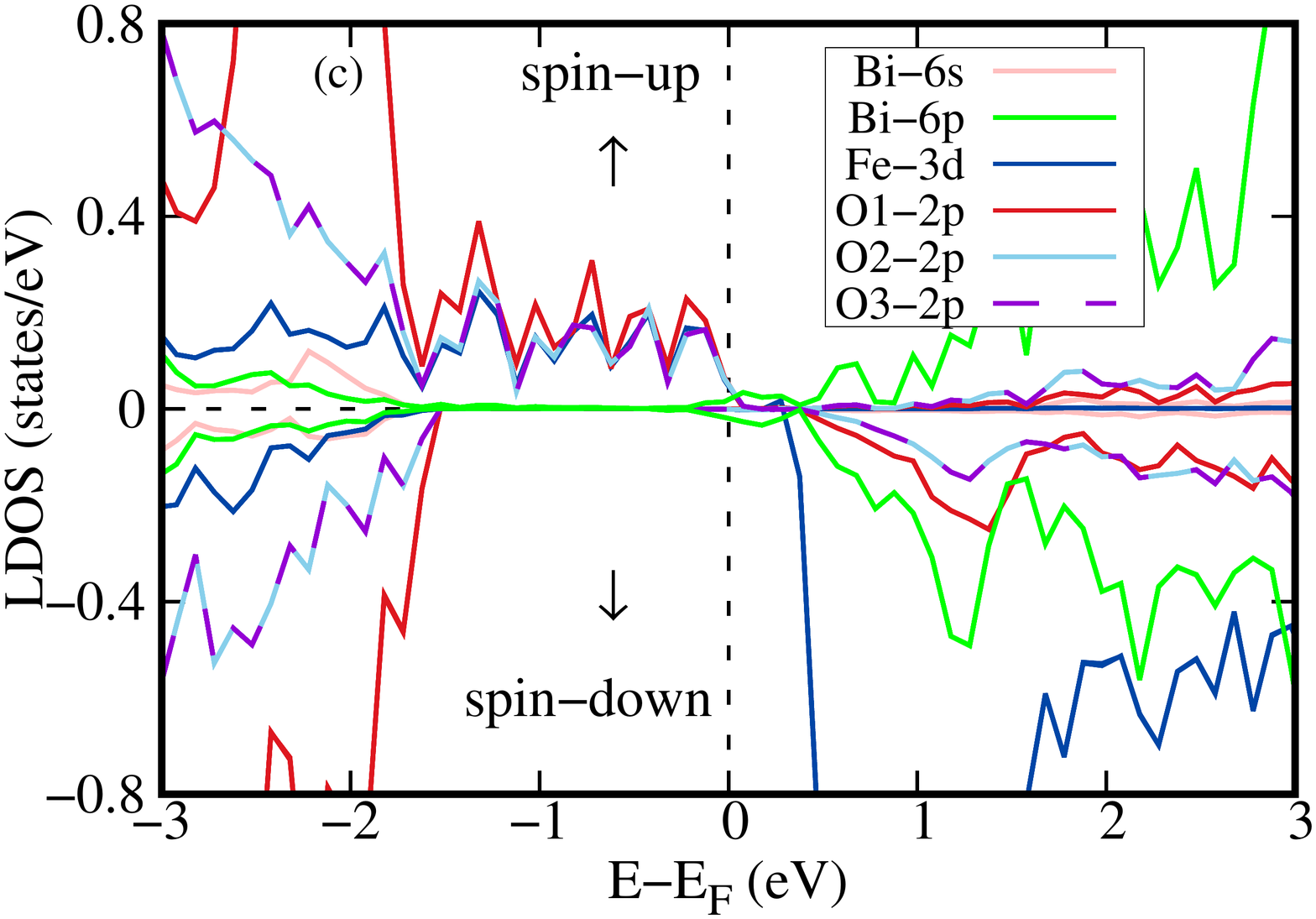}
	\vspace{-0.15cm}
\caption{The electronic properties of structure-III: (a) Band structure (BS), (b) Total density of states (TDOS), (c) Local density of states (LDOS) }
\label{Fig:fm-str2-1}
\end{figure*}
\begin{figure*}[htbp!]
	\includegraphics[width=0.28\textwidth,height=0.28\textwidth,trim=1.7cm 0.5cm 1.5cm 2.0cm,clip=true]{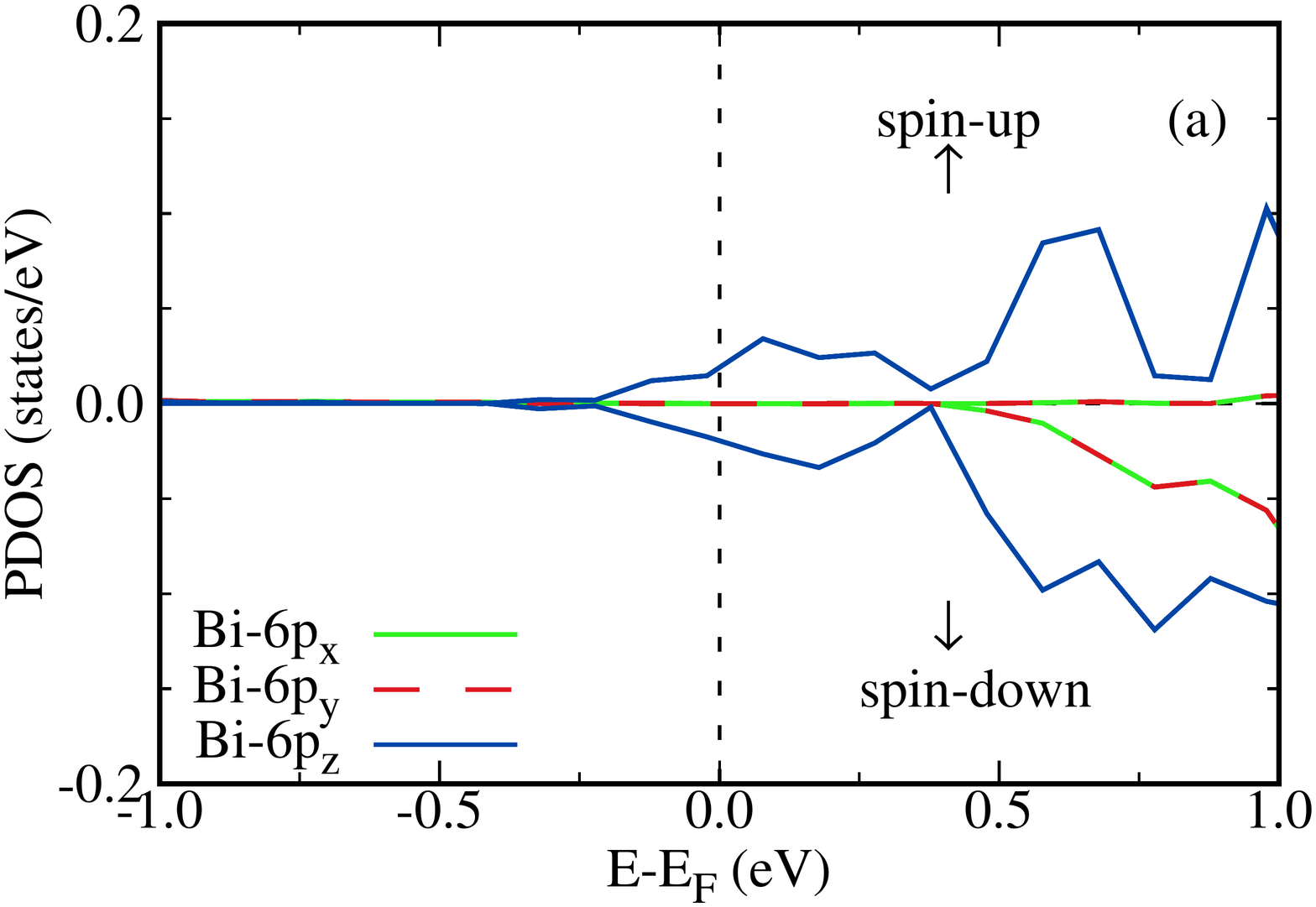}\hspace{-0.1cm}
	\includegraphics[width=0.28\textwidth,height=0.28\textwidth,trim=1.5cm 0.5cm 1.5cm 2.0cm,clip=true]{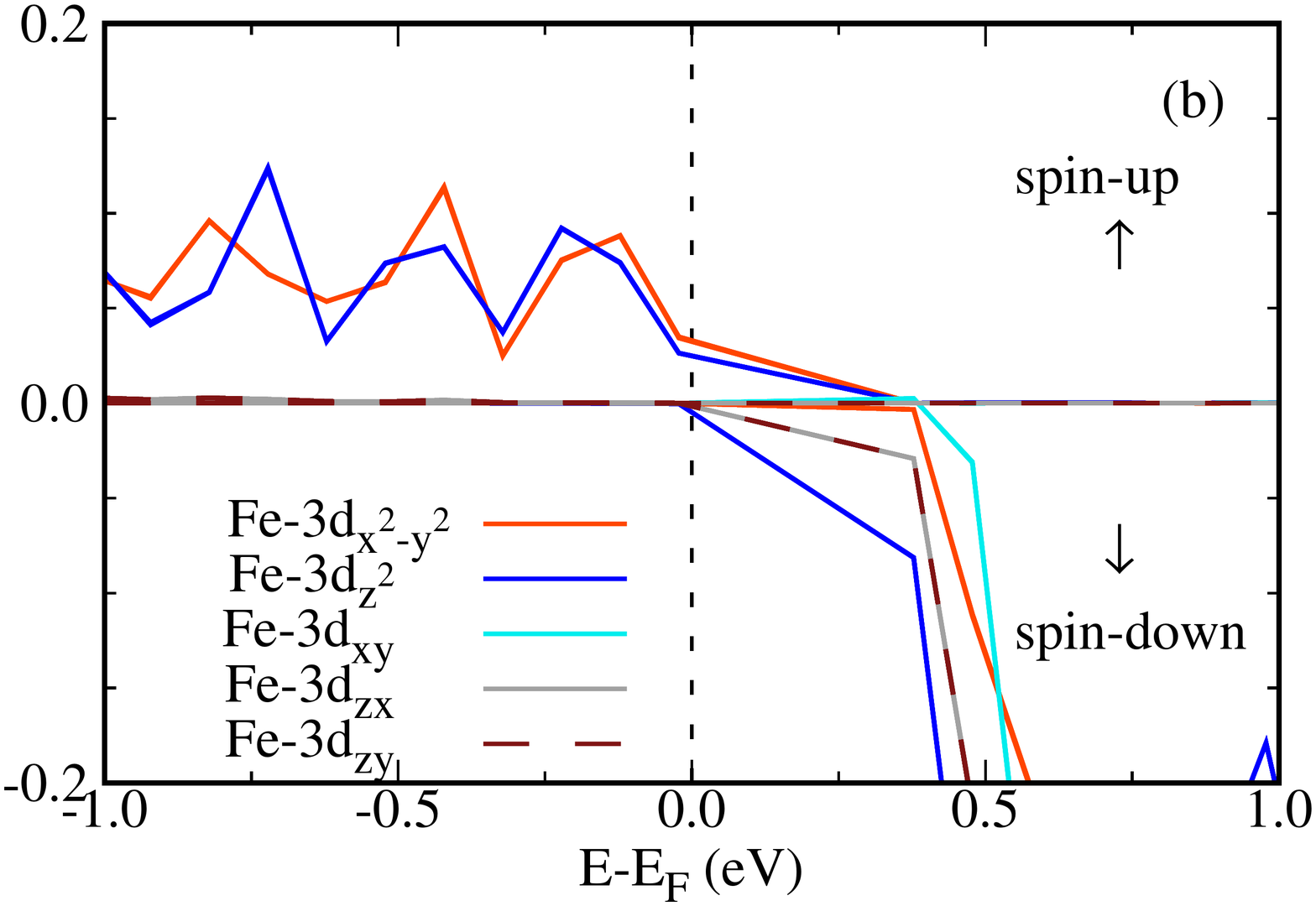}\hspace{-0.1cm}
	\includegraphics[width=0.28\textwidth,height=0.28\textwidth,trim=1.5cm 0.5cm 1.9cm 2.0cm,clip=true]{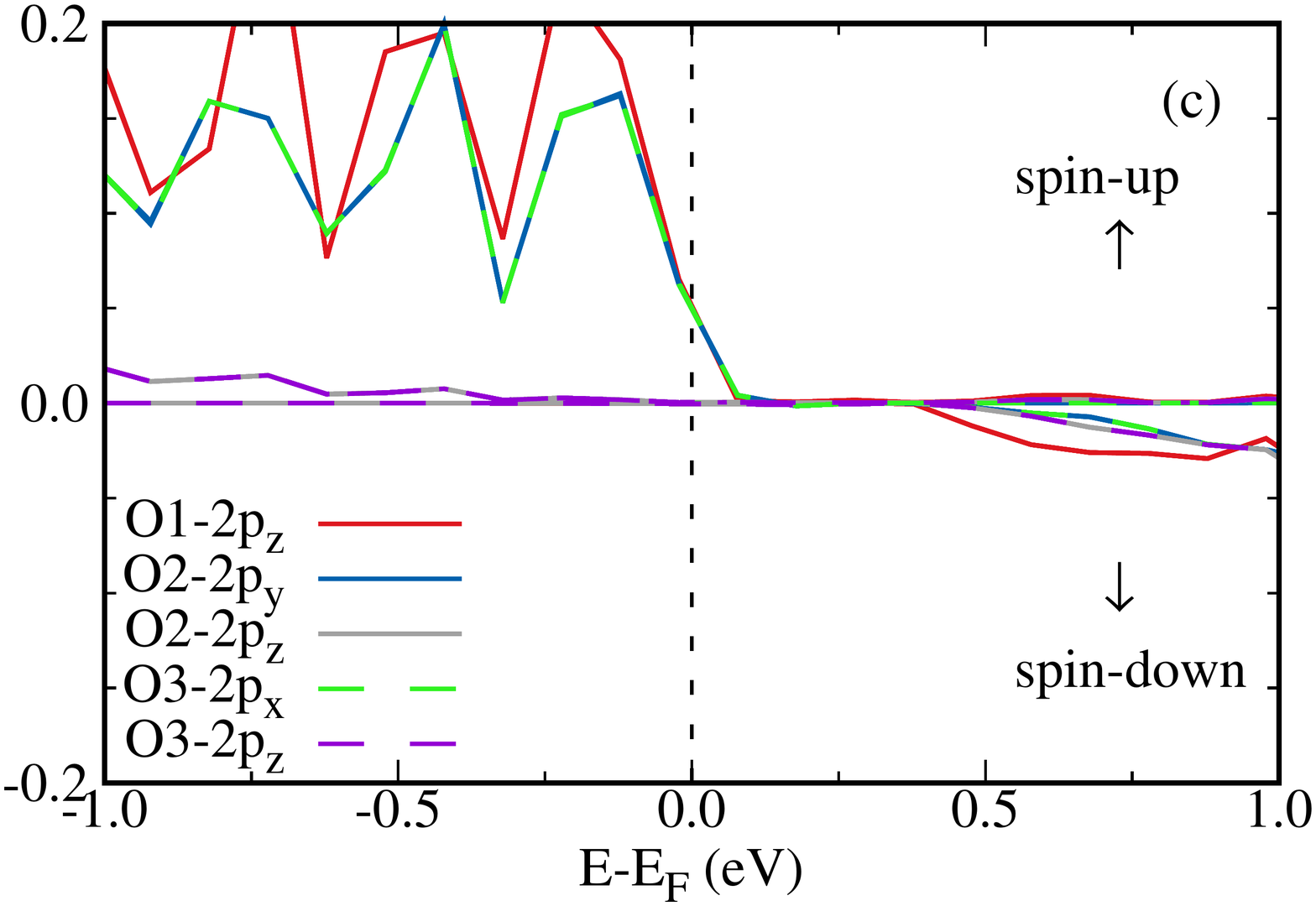}
\vspace{-0.6cm}
\caption{The projected density of states (PDOS) corresponding to the FM phase: (a) Bi-$6p_x,6p_y,6p_z$, 
(b) Fe-$e_g,t_{2g}$, (c) O1-$2p_z$, O2-$2p_y$, O3-$2p_x$ in structure-III. Here, the $\uparrow$ and $\downarrow$ 
belong to spin-up and spin-down component respectively.}
\end{figure*}\label{Fig:fm-str2-2}


\section{Results of structure-III}\label{Appendix:str3-bs}
After increasing the $c/a$ ratio from 1.016 to 1.049, we have found that structure-III also
shows purely metallic behavior, which is not very away from the properties of the structure-IV. 
Here, the band structure, TDOS, LDOS \& PDOS have been studied like all other structures. The 
metallicity in the band structure, i.e., Fig.\ref{Fig:fm-str2-1}(a) is clearly visible, which agrees with 
the corresponding TDOS (Fig.\ref{Fig:fm-str2-1}(b)). In the LDOS we found that, 
like structure-IV, Bi-$6p$, Fe-$3d$ and 
O-$2p$ have contributions near the Fermi energy. Bi-$6p$ contributes 
electron carriers from both the 
spin channels, while Fe-$3d$ and O-$2p$ orbitals contribute hole carriers to the system from the spin up channel only which is 
also in the case of structure-IV. 
In PDOS, we find that the Bi-$6p_z$ orbital contributes in both the spin-channels, whereas the $e_g$ band of `Fe' contributes along 
with the  O1-$2p_z$, O2-$2p_y$, and O3-$2p_x$ orbitals. These bands cross over the Fermi level for the spin-up 
channel only, thereby making the structure-III to be a ferromagnetic metal.

\section{L${\ddot{o}}$wdin charges}\label{Appendix:Lowdin-charge}
From L${\ddot{o}}$wdin charges analysis, it is found that the hybridization
of the principal central metal ion `Fe' are $sp^{2.420}d^{2.370}$ (structure-I),  $sp^{2.415}d^{2.366}$ (structure-II) and 
$sp^{2.414}d^{2.346}$ (structure-IV) respectively. So, the hybridization present in T-BFO can be 
called as the modified $s{p^2}{d^2}$. Also, we have found that the Bi-$5d$ does not have any contribution in the hybridization.  
The extra population of electrons in Fe-$3p$ \& Fe-$3d$ may be the outcome of the back-bonding, which has been created due to the 
interplay between Bi-$6s$ and Bi-$6p$.

\section*{ACKNOWLEDGEMENTS}
S. J. and S. D. would like to thank Prof. P. Sen for his generosity to access the high performance computing cluster 
facility at H.R.I, Prayagraj, India, where some of the initial calculations were performed. The authors are grateful 
to Prof. A. P. Chattopadhyay, University of Kalyani, West Bengal, India for very useful discussions. The authors 
would like to thank the computing facility provided by the National Institute of Technology, Rourkela and the 
computational facility provided by the Science and Engineering Research Board, Department of Science and Technology, 
India (Grant No: EMR/2015/001227).


\end{document}